\documentclass[draft]{agujournal2019}
\usepackage{url} 
\usepackage{lineno}
\usepackage[inline]{trackchanges} 
\usepackage{soul}
\usepackage{amsmath}
\usepackage{amsfonts}

\draftfalse

\journalname{Journal of Advances in Modeling Earth Systems (JAMES)}

\renewcommand{\vec}[1]{\boldsymbol{#1}}  
\newcommand{\refp}[1]{(\ref{#1})}  
\newcommand{\mat}[1]{\mathsf{#1}}

\newcommand{\trans}{\mathsf{T}}

\definecolor{rred}{rgb}{0.7,0,0.1}

\definecolor{gblue}{rgb}{0.0,0.7,0.6}

\definecolor{ggreen}{rgb}{0.1,0.7,0.1}

\begin{document}

%
%


\title{Identifying efficient ensemble perturbations for initializing subseasonal-to-seasonal prediction}

%
%




\authors{Jonathan Demaeyer\affil{1}, Stephen G. Penny\affil{2,3}, and St\'{e}phane Vannitsem\affil{1}}

\affiliation{1}{Royal Meteorological Institute of Belgium, Brussels, Belgium}
\affiliation{2}{Cooperative Institute for Research in Environmental Sciences, University of Colorado Boulder}
\affiliation{3}{NOAA Physical Sciences Laboratory, Boulder, Colorado}





\correspondingauthor{Jonathan Demaeyer}{jodemaey@meteo.be}




\begin{keypoints}
\item Several methods for initializing ensemble forecasts with long lead times are tested in the context of an ocean-atmosphere coupled model
\item The methods providing the most reliable ensembles are the adjoint Lyapunov vectors and the adjoint modes of the Dynamic Mode Decomposition
\item These are related to the eigenfunctions of the Koopman and Perron-Frobenius operators of the system
\end{keypoints}

%
%

%
%


\begin{abstract}
The prediction of the weather at subseasonal-to-seasonal (S2S) timescales is dependent on both initial and boundary conditions. An open question is how to best initialize a relatively small-sized ensemble of numerical model integrations to produce reliable forecasts at these timescales. Reliability in this case means that the statistical properties of the ensemble forecast are consistent with the actual uncertainties about the future state of the geophysical system under investigation. In the present work, a method is introduced to construct initial conditions that produce reliable ensemble forecasts by projecting onto the eigenfunctions of the Koopman or the Perron-Frobenius operators, which describe the time-evolution of observables and probability distributions of the system dynamics, respectively. These eigenfunctions can be approximated from data by using the Dynamic Mode Decomposition (DMD) algorithm. The effectiveness of this approach is illustrated in the framework of a low-order ocean-atmosphere model exhibiting multiple characteristic timescales, and is
compared to other ensemble initialization methods based on the Empirical Orthogonal Functions (EOFs) of the model trajectory and on the backward and covariant Lyapunov vectors of the model dynamics. Projecting initial conditions onto a subset of the Koopman or Perron-Frobenius eigenfunctions that are characterized by time scales with fast-decaying oscillations is found to produce highly reliable forecasts at all lead times investigated, ranging from one week to two months. Reliable forecasts are also obtained with the adjoint covariant Lyapunov vectors, which are the eigenfunctions of the Koopman operator in the tangent space. The advantages of these different methods are discussed.
\end{abstract}

\section*{Plain Language Summary}
Weather forecasts often reach their limit of predictability at one to two weeks. In order to extend forecast skill beyond this two week limit, the weather prediction community has begun transitioning to the use of coupled models that include both atmosphere and ocean dynamics, with the slower ocean dynamics enabling an extended forecast horizon. Due to uncertainties in the accuracy of the initial conditions and the model itself, such forecasts must be probabilistic. The primary approach for probabilistic weather prediction is to generate ensemble forecasts that integrate multiple copies of the model started from slightly different initial conditions. Here we show that the method used to determine the ensemble of initial conditions has a significant impact on the probabilistic forecast skill at horizons ranging from a few weeks to a few months. We show that many of the existing techniques used for short forecasts are suboptimal for longer forecast horizons. We introduce a new perspective and corresponding techniques that permit the initialization of these ensemble forecasts using information that is intrinsic to the nature of the evolution of the coupled system dynamics, and present data-driven methods that allow this information to be estimated directly from historical data.

\section{Introduction}
\label{sec:intro}

Long-term forecasts of the atmosphere at sub-seasonal, seasonal, and decadal time scales are affected by both the choice of the initial condition and the slow evolution of surface boundary conditions. This multi-timescale forecasting is a key target of the atmospheric and climate communities ~\cite{Vitart2017,Cassou2018}. Forecast error grows
quickly at increasing lead times due to the instability properties of weather dynamics. For this reason, a probabilistic approach is necessary in order to isolate the multiple possible outcomes of a set of forecasts. Since the 1990's, such an approach has been developed in many operational weather prediction centers based on using multiple numerical integrations of the models starting from slightly different initial conditions. 
This approach is known as ensemble forecasting~\cite{Buizza2019, Kalnay2019}. Different perturbation techniques have been designed for initializing ensemble weather forecasts, of which the most popular are the singular vectors~\cite{Molteni1996}, the bred modes~\cite{Toth1997}, and perturbed observations applied within data assimilation systems~\cite{Buizza2005, kleist2015osse}.
\citeA{Buizza2005} noted that the accuracy of initial conditions is just as important as the accuracy of the forecast models for generating reliable ensemble forecasts.

As operational centers expanded their focus to include longer forecast horizons, the same perturbation approaches were also used for sub-seasonal to decadal predictions. However, at timescales beyond the limits of predictability for the atmosphere, coupled Earth system models must be used. This introduces the additional difficulty of building appropriate perturbations for the different components of such multi-scale systems~\cite{OKane2019}. Several approaches consisting of appropriately tuning the bred modes to capture the long time scales of the dynamics have for instance been proposed~\cite{Pena2004, Yang2008, OKane2019}, and the use of backward Lyapunov vectors (BLV), closely related to the bred modes, have been used to build reliable ensemble forecasts in idealized scenarios~\cite{VD2020}.   

In the present work, we address this ensemble initialization problem by considering tools coming from the probabilistic description of dynamical systems and finding their roots in the conservation of the number of trajectories in phase space described by the Liouville equation~\cite{G2005,NN2012}. The evolution operator associated with this equation is known as the Perron-Frobenius operator~\cite{LM2008}, sometimes also called the \emph{transfer} operator. It has been used as a theoretical framework to describe \emph{probabilistic forecasting}~\cite{E2006, G2019}, i.e. forecasting based on the time evolution of a probability distribution, and for which the ensemble forecasting methods provide approximations.
The adjoint of the Perron-Frobenius operator, known as the Koopman operator, has become popular to describe the dynamics of observables on attractors~\cite{mezic2013,susuki2016,AM2017,santos2021}, due to the fact that when operating on functional spaces it is a linear operator, an \emph{observable} being defined as any function mapping the system state to some real or complex value. A trade-off, however, in converting the nonlinear dynamics to a linear representation is that the Koopman operator generally acts on an infinite dimensional space, but as we will see, methods exist to obtain finite-dimensional approximate representations of these operators. The computation of the spectrum of these operators has also been considered in order to study bifurcations in low- and high-dimensional systems~\cite{TLD2018,TVLD2018}.
The eigenvalues and eigenfunctions of these operators can then be obtained in the functional spaces, and provide the key building blocks of the dynamics of the probability density and observables. These are precisely the quantities that are used in the present work to generate the ensemble forecasts initialization, as they constitute generic features of the dynamics of the probability density.

The eigenfunctions of the Koopman operator can be approximated using Dynamic Mode Decomposition (DMD)~\cite{RMBSH2009, TRLBK2014}. The DMD approach is a rediscovery of the Linear Inverse Model (LIM), which was developed first within the seasonal prediction community~\cite{P1989, Penland1993, penland1995}. For computational efficiency, and due to the large volume of data involved, LIMs are typically formed using data projected to the space of Empirical Orthogonal Functions (EOFs) and then truncated. As such, a mathematical equivalence between this form of the LIM and the projected DMD was noted by~\citeA{TRLBK2014}. The LIM approach is now being used experimentally for seasonal forecasts by the US National Oceanographic and Atmospheric Administration (NOAA) Climate Prediction Center (CPC) \cite{w24-NOAACPCLIM-website}. Since its rediscovery by the fluid mechanics community under the name of DMD~\cite{S2010}, many new algorithms, theoretical results, and variants have been developed that have advanced understanding of this approach. Most notably, perhaps, is the connection between DMD and the Koopman operator~\cite{RMBSH2009}.

The usefulness of the Perron-Frobenius and Koopman operators for producing ensemble forecasts will be analyzed in a reduced order coupled ocean-atmosphere model, previously demonstrated for a similar purpose by~\citeA{VD2020}. The model will be briefly described in Section \ref{sec:OAmodel}. The experimental setup will be then presented in Section \ref{sec:expdesign}.  
In Section \ref{sec:initmethods}, the different bases onto which the perturbed initial conditions are projected will be presented: first, the EOFs that are often used in initializing climate models and their ensemble integrations, e.g.~\citeA{Polkova2019}; second, the Lyapunov vector approach used by~\citeA{VD2020}, which is closely related to the bred modes and ensemble Kalman filters; and third, the eigenfunctions of the Koopman and Perron-Frobenius operators. The Koopman and Perron-Frobenius operators are extensively discussed, as important clarifications on their link with DMD is needed. Section \ref{sec:bases} describes the specific choices of bases used for the experiments. Section \ref{sec:results} presents the results of experiments using the aforementioned bases to initialize ensemble forecasts. It will be shown that the eigenfunctions of the Koopman and Perron-Frobenius operators are indeed the most efficient tools for producing reliable ensemble forecasts in such multiscale systems.
Finally, conclusions are drawn in Section \ref{sec:conclusions}.

\section{The coupled ocean-atmosphere model}
\label{sec:OAmodel}

Experiments are conducted with a coupled ocean--atmosphere model that was first introduced by~\citeA{VDDG2015}, and was further generalized by~\citeA{DDV2016} and~\citeA{DDV2020}. It consists of a two-layer quasi-geostrophic atmospheric model coupled both thermally and mechanically to a shallow-water oceanic component on a beta plane. The coupling between the ocean and the atmosphere includes the wind stress and heat exchanges. The fields of the model are defined on a rectangular domain with the zonal and meridional coordinates $x$ and $y$ being restricted to $0 \leq x \leq 2 \pi L/n$ and $0 \leq y \leq \pi L$, where $n$ is the aspect ratio of the domain and $L$ is the characteristic spatial scale. The atmospheric fields are defined in a zonally periodic channel with no-flux boundary conditions in the meridional direction, i.e. if $\psi$ is such an atmospheric field then $\partial \psi /\partial x \equiv 0$ at $y = 0, \pi L$. The oceanic fields are defined on a closed basin, with no flux through the boundaries. 

The model fields include the atmospheric barotropic $\psi_{\rm a}$ and baroclinic streamfunctions $\theta_{\rm a}$, and the ocean streamfunction $\psi_{\rm o}$ and the temperature field $\theta_{\rm o}$. These fields are expanded in series of Fourier modes $F_i(x,y)$ for the atmosphere and $\phi_i(x,y)$ for the ocean, both respecting the prescribed boundary conditions:
\begin{align*}
    \psi_{\rm a}(x,y) & = \sum_{i=1}^{n_{\rm a}} \, \psi_{{\rm a}, i} \, F_i(x,y) \\
    \theta_{\rm a}(x,y) & = \sum_{i=1}^{n_{\rm a}} \, \theta_{{\rm a}, i} \, F_i(x,y) \\
    \psi_{\rm o}(x,y) & = \sum_{i=1}^{n_{\rm o}} \, \psi_{{\rm o}, i} \, \phi_i(x,y) \\
    \theta_{\rm o}(x,y) & = \sum_{i=1}^{n_{\rm o}} \, \theta_{{\rm o}, i} \, \phi_i(x,y)
\end{align*}
After projecting the partial differential equations (PDEs) of the model on the Fourier modes, one obtains a set of ordinary differential equations (ODEs) governing the time evolution of the coefficients $\psi_{{\rm a}, i}$, $\theta_{{\rm a}, i}$, $\psi_{{\rm o}, i}$ and $\theta_{{\rm o}, i}$:
\begin{align}
    \dot{\vec{x}} & = \vec f(\vec x) \label{eq:sysdyn} \\ 
    \vec x & = [\psi_{{\rm a}, 1}, \ldots, \psi_{{\rm a}, n_{\rm a}}, \theta_{{\rm a}, 1}, \ldots, \theta_{{\rm a}, n_{\rm a}}, \nonumber \\
    & \qquad \qquad \psi_{{\rm o}, 1}, \ldots, \psi_{{\rm o}, n_{\rm o}}, \theta_{{\rm o}, 1}, \ldots, \theta_{{\rm o}, n_{\rm o}}]^\trans \nonumber
\end{align}
where $^\trans$ denotes the matrix transposition operation.
These coefficients thus form the set of the model state variables and the equation above allows one to simulate the physical system using numerical integration. In the present study, we consider the so-called \emph{VDDG} model configuration first defined by~\citeA{VDDG2015}, with the atmospheric and the oceanic fields each being expanded into a series of $n_{\rm a}=10$ and $n_{\rm o}=8$ selected modes, respectively, leading to a system with $d=36$ dimensions\footnote{In the following, the letter $d$ will always refer to the dimension of the dynamical system.}.

\begin{figure*}
\includegraphics[width=0.495\textwidth]{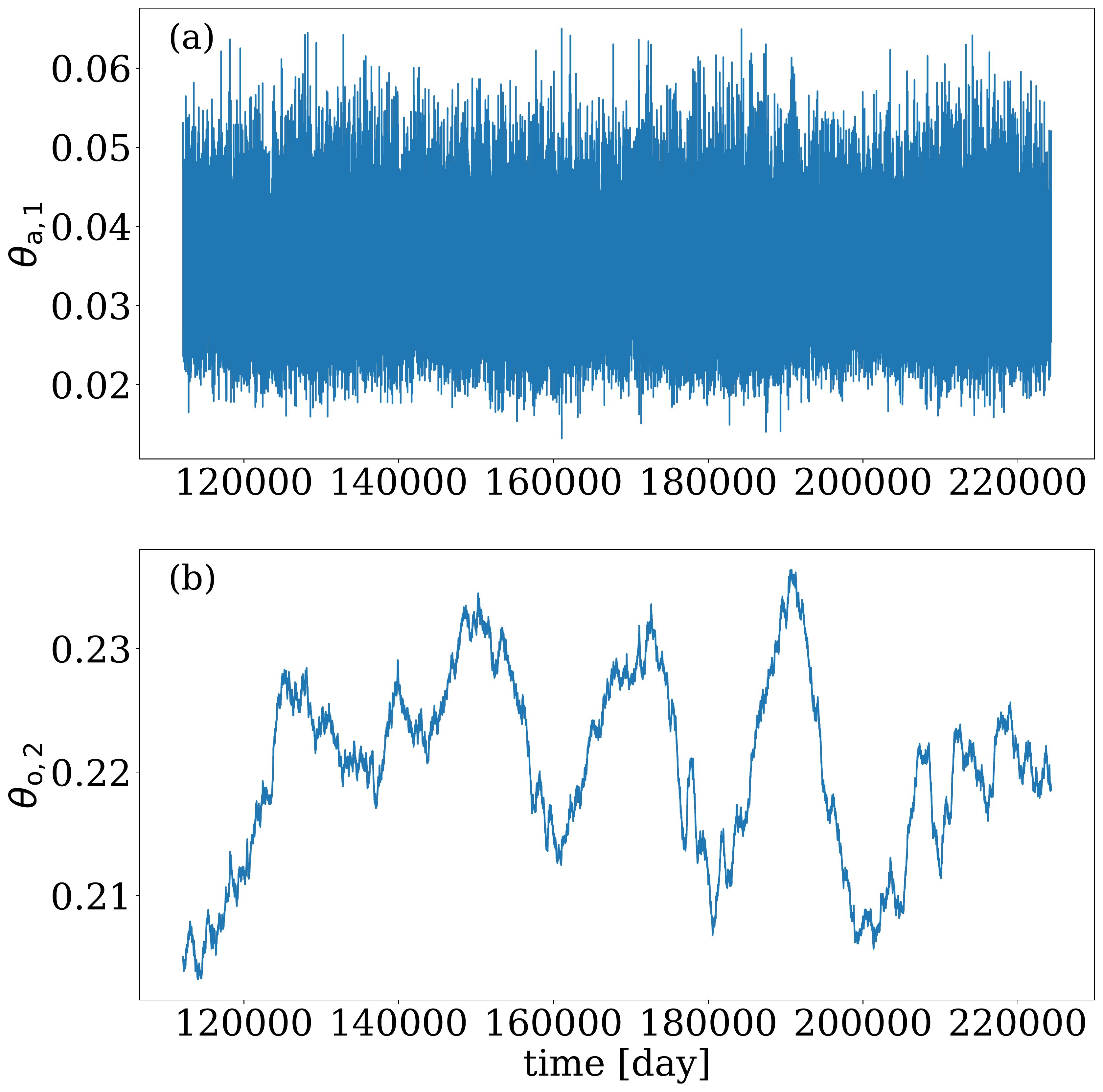}
\includegraphics[width=0.495\textwidth]{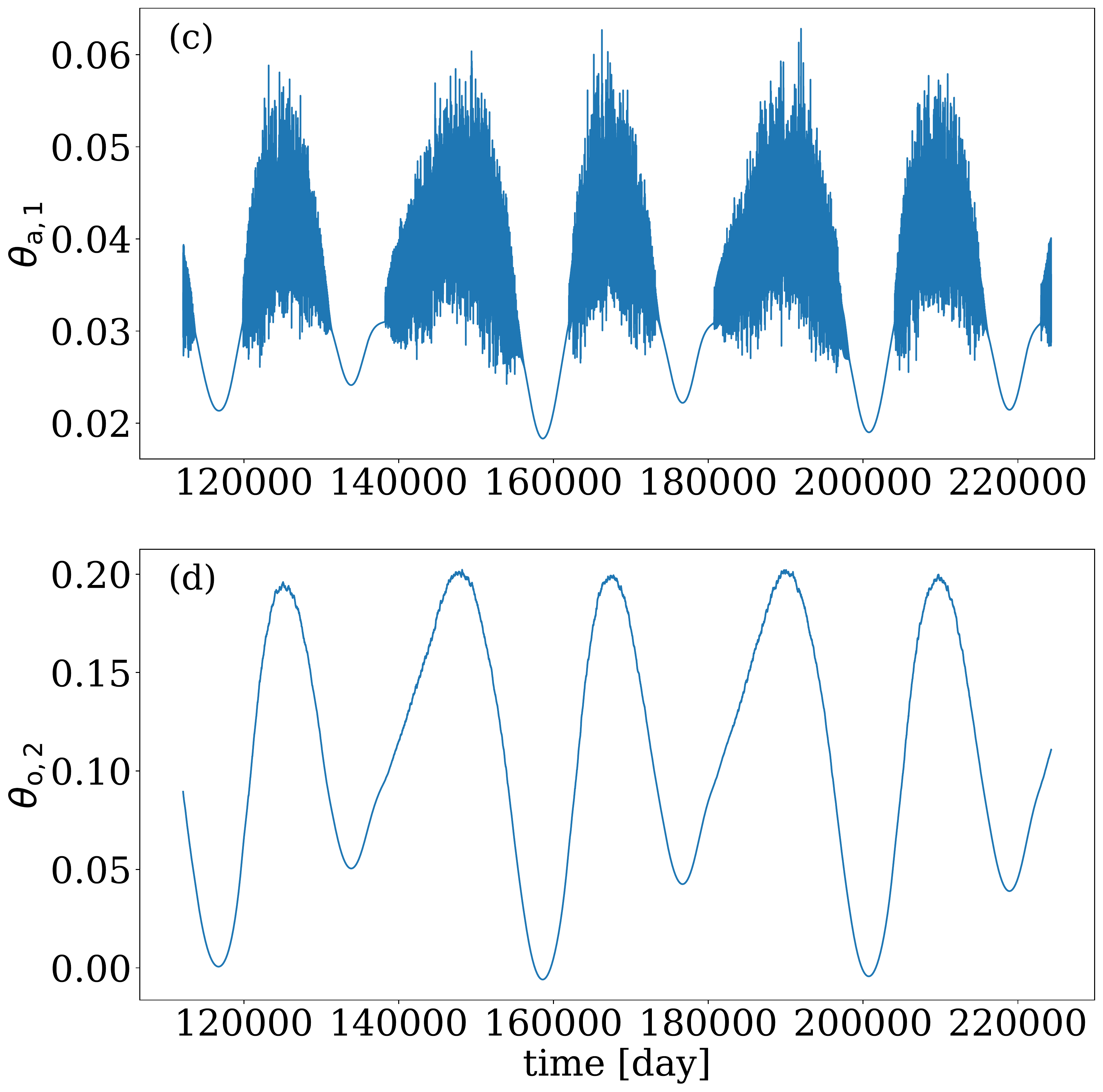}
\caption{Time evolution of a trajectory on the attractor for: (a),(b) the case without low-frequency variability and (c),(d) the case with low-frequency variability. For this latter case, it shows that the presence or absence of atmospheric instability and fast variability is linked to the low-frequency variation of the meridional temperature gradient in the ocean.}
\label{fig:timevol}
\end{figure*}

A critical parameter of the model is the friction coefficient $C$ between the ocean and the atmosphere. Indeed, it was shown by~\citeA{VDDG2015} that the strength of the wind stress controls the presence and the amplitude of a low-frequency variability (LFV) typically found in the real atmosphere at midlatitude. Following~\citeA{V2017} and~\citeA{VD2020}, we shall consider two cases: one with weak LFV ($C=0.01$ kg m$^{-2}$ s$^{-1}$) and another with much more pronounced LFV ($C=0.016$ kg m$^{-2}$ s$^{-1}$). Solutions of the models for both cases are depicted in Figure~\ref{fig:timevol}, where the difference in the amplitude of LFV between the left and right panels is clear. The variables shown in this figure are the coefficients corresponding to the first mode of the baroclinic streamfunction and the second mode of the ocean temperature field, each sampled every $\Delta t = 10$ nondimensional model timeunits (MTU), corresponding to $1.1215$ days. The former mode is related to the meridional temperature gradient in the system, while the second corresponds to a dominant double-gyre signal in the ocean.

These two cases will allow us to highlight how the different methods of initialization that we consider perform in different settings, with different timescales and different correlation structures between the components being involved.

\section{Experiment design}
\label{sec:expdesign}
The focus of this study is initialization methods for ensemble forecasts. To this end, the long reference runs depicted in Figure~\ref{fig:timevol} were computed to serve as the `truth' in our experiments. We select $N$ points of the reference runs, denoted $\vec x_n(0)$, as initial conditions to produce $N$ ensemble forecasts with the \emph{VDDG} model, using the same parameters as the reference runs.
To ensure that the experiments are initialized from a state close to the true trajectory, but with an ensemble mean state that is not precisely equal to the truth, we first obtain the initial conditions $\vec x_n^{\rm ctrl}$ of a deterministic control forecast by perturbing the $N$ points of the reference `truth' by a random perturbation $\vec{\delta x}^{\rm ctrl}_0$ sampled from a uniform distribution $\rho_0^{\rm pert}$:
\begin{equation}
    \vec x_n^{\rm ctrl}(0) = \vec x_n(0) + \vec{\delta x}^{\rm ctrl}_0.
\end{equation}
An ensemble is then generated by perturbing the control run initial conditions with a set of $M-1$ perturbations $\vec{\delta x}^m_0$ drawn from a distribution $\tilde\rho_0^{\rm pert}$, which is taken to be the same distribution as that used to obtain the control: $\tilde\rho^{\rm pert}_0 \equiv \rho^{\rm pert}_0$. The initial conditions of the ensemble are thus:
\begin{equation}
    \label{eq:IC}
    \vec y_{m,n}(0) = \vec x_n^{\rm ctrl}(0) + \vec{\delta x}^m_0 \quad , \quad m = 1,\ldots,M-1.
\end{equation}
Including the control run,  i.e. $\vec{\delta x}^M_0 = 0$, this provides a reference \emph{perfect} ensemble of $M$ members. In the present study, as in~\citeA{VD2020}, a uniform distribution defined on the interval $[-\varepsilon/2, \varepsilon/2]$ with $\varepsilon=10^{-6}$ was used to perturb each component of the system's state vector. Additional computations done with Gaussian distributions with the same standard deviation did not show any differences in the results of the study.

Due to the high dimensionality of more realistic applications, and the cost involved in integrating long model forecasts of this size, ensemble perturbations must be built from a relatively small subspace of the original system. While this distribution could be sampled randomly, we seek a set of initial conditions that can efficiently reproduce the true error growth characteristics {of the coupled system dynamics}.
The question remains as to what bases are most efficient for initializing a reliable ensemble forecast, and how can those bases be practically constructed in a realistic setting.

We examine reduced-size ensembles constructed using linear projections of the `perfect' ensemble initial conditions onto various bases forming subspaces with rank less than $M$, and compare these to the perfect ensemble as a benchmark. The specific bases that we use will be detailed in the next section. By design, the ensemble perturbations obtained by projection cannot be more reliable than the original reference ensemble. However, we will show that depending on the modes and subspaces selected as a basis, these reduced-size ensembles can achieve similar performance to the full-rank `perfect' ensemble. 

To determine whether the ensemble forecasts generated from the projected initial conditions are reliable, the mean square error (MSE) of the ensemble mean and the variance of the ensemble (the square of the ensemble spread) are computed at each lead time $\tau$ of the ensemble forecasts as:
\begin{align}
    \mathrm{MSE}(\tau) & = \frac{1}{N} \sum_{n=1}^N \, \left\| \vec{x}_n(\tau) - \bar{\vec{y}}_n(\tau)\right\|^2 \label{eq:MSE}  \\
    \mathrm{Spread}^2(\tau) & = \frac{1}{N} \sum_{n=1}^N \frac{1}{M-1} \sum_{m=1}^M \, \left\| \vec{y}_{m,n}(\tau) - \bar{\vec{y}}_n(\tau)\right\|^2 \label{eq:Spread}
\end{align}
where
\begin{equation}
    \label{eq:EM}
    \bar{\vec{y}}_n(\tau) = \frac{1}{M} \sum_{m=1}^M \, \vec{y}_{m,n}(\tau) 
\end{equation}
is the ensemble mean over the members $\vec{y}_{m,n}(\tau)$ of the $n^{\mathrm{th}}$ ensemble forecast and $\vec{x}_n(\tau)$ is the corresponding reference solution. Finally, $\|\cdot\|^2$ is the L$^2$-norm. If the $\mathrm{Spread}^2$ and the $\mathrm{MSE}$ are close to one another, indicating that the estimated error is close to the true error, then the ensemble forecast is considered reliable~\cite{LP2008}. The results based on these measures are presented in the supplementary material.

An alternative measure of reliability of the ensemble forecasts can also be assessed by considering the proper ignorance (or logarithmic) score~\cite{RS2002}:
\begin{equation}
    \mathcal{I}[\rho^{\rm ens}_\tau] = - \ln \rho^{\rm ens}_\tau (\vec x_n(\tau) | \vec x^{\rm ctrl}_n(0)) .
\end{equation}
Applying the ignorance score to a Gaussian, one obtains the related proper two-moment skill score derived by~\citeA{DS1999}. As such, and regardless of whether the distributions being considered is Gaussian or not~\cite{L2019}, the Dawid-Sebastiani Score (DSS) provides an evaluation of the quality of the first and second moments of the forecast distribution estimated by the ensemble, with respect to the true moments. The bias-free univariate DSS for the $n^{\mathrm{th}}$ ensemble forecast and the $i^{\mathrm{th}}$ variable of the system can be written as~\cite{SFSL2019}:
\begin{align}
    \mathrm{DSS}_{n,i}(\tau) = & \frac{1}{2} \, \log(2\pi) + \frac{1}{2} \, \log \, \sigma_{n,i}^2(\tau) \nonumber \\
    & + \left.\frac{1}{2} \frac{M-3}{M-1} \, \left( \bar{y}_{n,i}(\tau) - x_{n,i}(\tau)\right)^2 \right/ \sigma_{n,i}^2(\tau), \label{eq:DSS}
\end{align}
where $\sigma_{n,i}^2$ is an estimator of the $i^{\mathrm{th}}$ variable ensemble variance:
\begin{equation}
    \sigma_{n,i}^2(\tau) = \frac{1}{M-1} \sum_{m=1}^M \, | y_{m,n,i}(\tau) - \bar{y}_{n,i}(\tau)|^2.
\end{equation}
This score can then be averaged over the $N$ realizations:
\begin{equation}
    \mathrm{DSS}_i(\tau) = \frac{1}{N} \sum_{n=1}^N \, \mathrm{DSS}_{n,i}(\tau).
\end{equation}
The lower the DSS score, the more reliable the ensemble forecasts are for this particular variable. In particular, the DSS score has been used to characterize the ensembles reliability in the study done by~\citeA{VD2020}.

\section{Initialization methods for ensemble forecasts}
\label{sec:initmethods}

We now discuss the different bases onto which the set of ensemble perturbations will be projected. Assume that a basis comprises $k$ vectors of dimension $d$ arranged as columns of the matrix $\mat B \in \mathbb{C}^{d\times k}$. We can construct the projection operator onto this basis as,
\begin{equation}
    \label{eq:gprojector}
    \mat\Pi = \mat B ( \mat B^\ast \, \mat B )^{-1} \, \mat B^\ast.
\end{equation}
If $\mat B$ is unitary, this reduces to $\mat\Pi = \mat B \mat B^\ast$~\cite{M2000}. Assuming that $\mat\Pi \in \mathbb{R}^{d \times d}$, if one considers the ensemble of $M-1$ perturbations $\vec{\delta x}^m_0$ of the control initial conditions, then the projection of the perturbations onto the subspace spanned by $\mat B$ is given by:
\begin{equation}
    \vec{\delta x}'^{m}_{\, 0} = \mat\Pi \, \vec{\delta x}^m_0.
\end{equation}
The resulting perturbations are used to initialize ensemble forecasts with the initial conditions:
\begin{equation}
    \label{eq:pertIC}
    \vec y_{m,n}'(0) = \vec x_n^{\rm ctrl}(0) + \vec{\delta x}'^{m}_{\, 0} \quad , \quad m = 1,\ldots,M
\end{equation}
in the experiments discussed in Section \ref{sec:results}. 

Let us now detail the various basis vectors considered and the subspaces that they span, namely the EOFs, the backward (BLVs) and covariant (CLVs) Lyapunov vectors, and the Koopman and Perron-Frobenius eigenfunctions determined using DMD. Unlike the other basis vectors used, the Lyapunov vectors are time-dependent, defined locally at each point of the reference trajectory, and are related to the stability of the local linearized dynamics.

\subsection{Empirical Orthogonal Functions}
\label{sec:EOFs}
The EOFs of the dataset are obtained using a Principal Component Analysis (PCA), which decomposes the data into a set of orthogonal basis functions and time-dependent coefficients. These orthogonal patterns can be obtained directly by singular value decomposition (SVD) of the data matrix, or by computing the eigenvectors of the data covariance matrix~\cite{Wilks2011}.

Assuming that the dataset is represented by the matrix $\mat X = [\vec x_0 \ldots \vec x_{K-1} ]$, with $\mat X \in \mathbb{R}^{d\times K}$. The columns of $\mat X$ are the system states $\vec x_k = \vec\Phi^{t_k}(\vec x_0)$ at times $t_k = k \Delta t$ where $\vec\Phi^t$ is the flow of the system~\refp{eq:sysdyn}: $\vec x(t) = \vec\Phi^{t} (\vec x(0))$, The EOFs are the column vectors of $\mat{U}$ as determined by the PCA:
\begin{equation}
    \mat T =  \mat U^\ast  \bar{\mat{X}}
\end{equation}
where $ \bar{\mat{X}} = \mat X - \langle \mat X \rangle_k$ is the matrix of system states with zero empirical time mean, and $\mat U$ is a matrix whose columns are the orthogonal eigenvectors of the matrix $ \bar{\mat{X}}  \bar{\mat{X}}^\ast$ which is proportional to the covariance matrix of the system, and $\mat T$ is the time-series of the coefficients of the decomposition. The eigenvalues of the matrix $ \bar{\mat{X}}  \bar{\mat{X}}^\ast$ are related to the variance of the data projected onto the corresponding mode. The amplitude of the eigenvalues comparatively to the others then provide the `fraction of explained variance' by a given EOF.

The EOFs can alternatively be obtained by SVD of $ \bar{\mat{X}}$:
\begin{equation}
     \bar{\mat{X}} = \mat U \mat\Sigma \mat V^\ast
\end{equation}
where $\mat U$ and $\mat V$ are two unitary square matrices and $\mat\Sigma$ is diagonal, containing the singular values of $ \bar{\mat{X}}$. The matrix $\mat U$ contains the EOFs of $ \bar{\mat{X}}$ since $ \bar{\mat{X}}  \bar{\mat{X}}^\ast = \mat U \mat\Sigma \mat\Sigma^\ast \mat U^\ast$, and the PCA time-series of coefficients can be represented as $\mat T = \mat U^\ast  \bar{\mat{X}}  = \mat \Sigma \mat V^\ast$.

\subsection{The Lyapunov Vectors}
\label{sec:BFCLVs}

We next consider the backward Lyapunov vectors (BLVs), the covariant Lyapunov vectors (CLVs), and the adjoint CLVs as basis vectors $\mat B$ in Eq.~\refp{eq:gprojector}. The Lyapunov vectors are locally defined in the tangent space of the trajectory of the model, and give information about the stability therein. For instance, Osedelets has shown that the tangent space can be decomposed into a set of nested subspaces $S^-_k$ that are invariant under the tangent linear model dynamics~\cite{O1968,O2008}. Arbitrary $k$-volumes defined in the tangent space converge to the subspace $S^-_k$ under the action of the tangent flow. These subspaces are spanned by the BLVs $\vec\varphi^-_i$: $S^-_k  = \mathrm{Span}\{\vec\varphi^-_i | i=1,\ldots,k\}$. The BLVs are thus related to the asymptotic properties of volumes in the tangent space, i.e. to how volumes contract or expand in the tangent space. The Lyapunov exponents characterize the time-average expansion and contraction rates of these volumes over the entire attractor. 

The CLVs $\vec{\varphi}_i$ are defined as stability directions in the tangent space that are covariant under the application of the tangent linear model dynamics. The tangent linear flow maps a CLV at one time to the same CLV at a later time, but multiplied by a \emph{stretching factor} defined over the same timescale as the tangent linear mapping, which indicates the local stability of this CLV~\cite{KP2012}. Finally, the adjoint CLVs $\tilde{\vec{\varphi}}_i$ are vectors that are covariant as well, but with respect to the flow of the adjoint model. See~\ref{sec:appLyap} for more details.

\begin{figure*}
\includegraphics[width=0.495\textwidth]{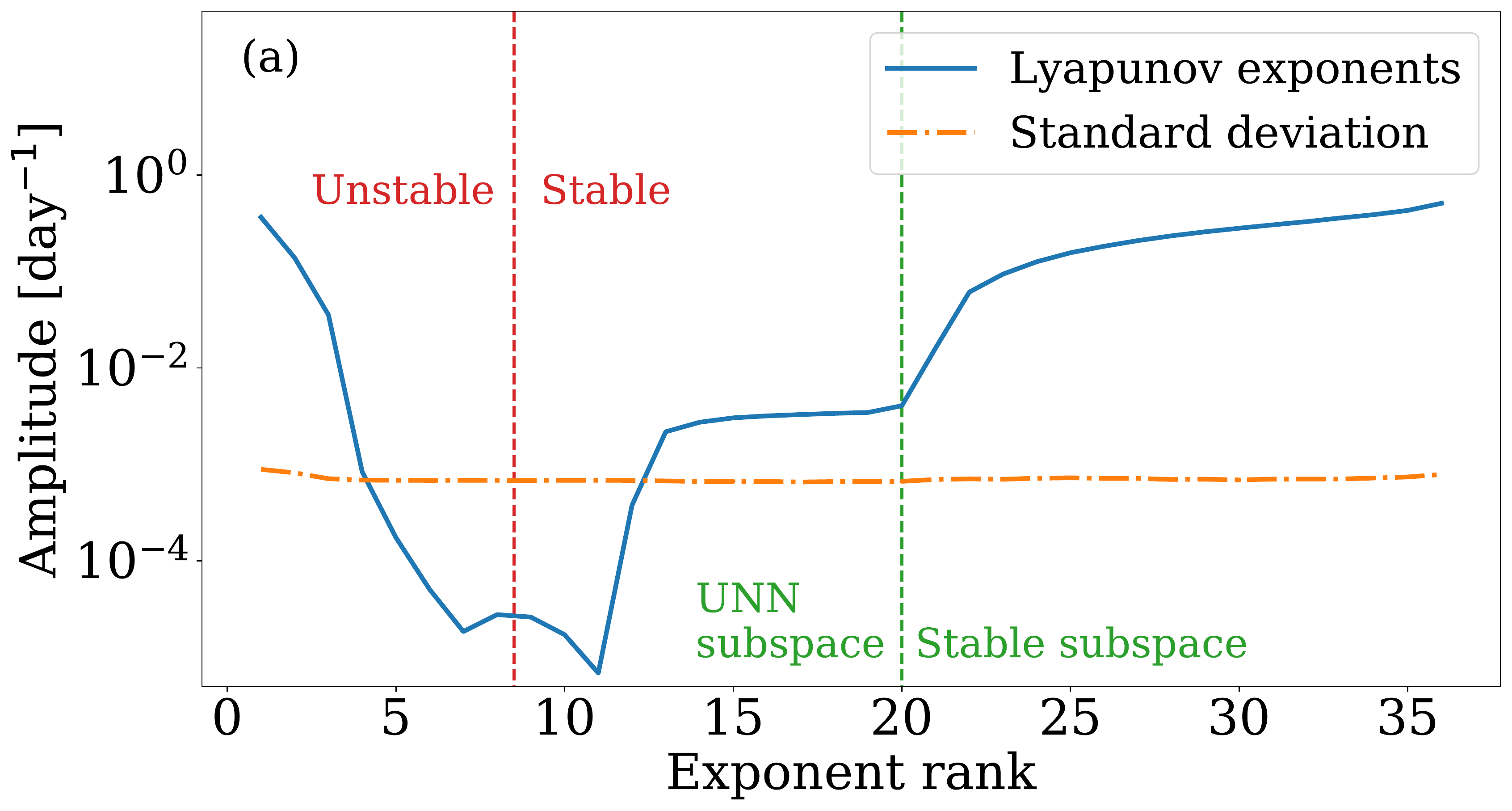}
\includegraphics[width=0.495\textwidth]{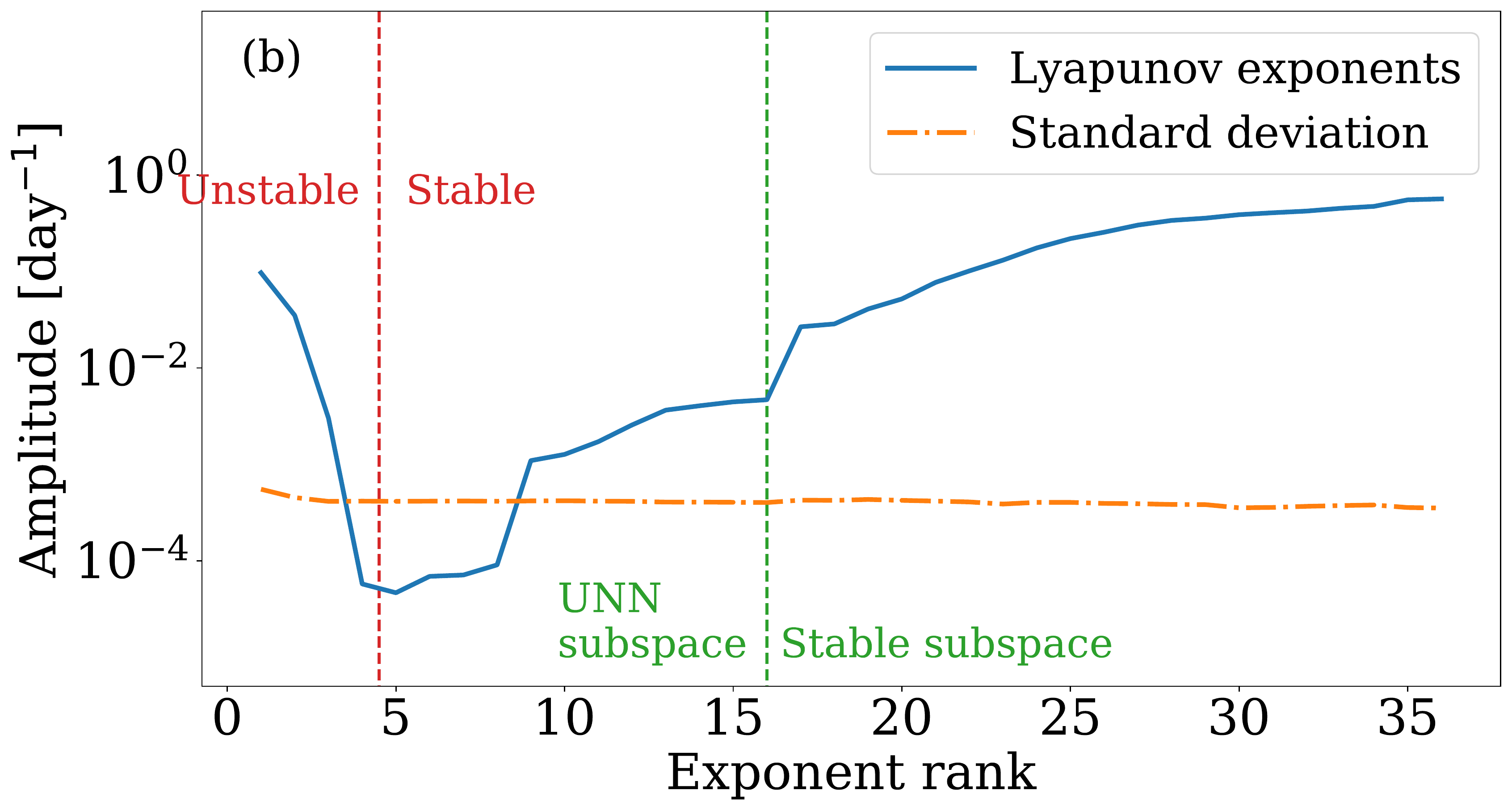}
\caption{Absolute value of the Lyapunov exponents (the LEs are expressed in $\mathrm{day}^{-1}$), along with the one standard-deviation uncertainty,  for: (a) the case with weak low-frequency variability and (b) the case with strong low-frequency variability. The approximate} separation between the positive (unstable) and negative (stable) exponents is depicted by a vertical dashed red line.
\label{fig:Lyapunov}
\end{figure*}

To determine the sets $\mat B$ of basis vectors that we will consider in the experiments, it is useful to consider the Lyapunov spectra $\sigma_i$ (depicted in Figure~\ref{fig:Lyapunov}).
These exponents have been estimated by averaging the \emph{local stretching rate} \footnote{See Eq.~\refp{eq:lyapexp} in~\ref{sec:appLyap}.} $\chi_i$ along the trajectories depicted in Figure~\ref{fig:timevol} with a bootstrap algorithm~\cite{Efron1993} to increase its statistical significance. The standard deviation of the time series used to compute the averaged Lyapunov exponent is also shown.

A chaotic dynamical system generally has positive (unstable) and negative (stable) exponents, along with a single zero-valued exponent that corresponds to the direction of flow of the system trajectory. For the coupled atmosphere-ocean system, however, because the magnitude of many of the near-zero exponents is smaller than the standard deviation of the time series itself, it is difficult to precisely identify the zero-valued Lyapunov exponent that separates the stable and unstable directions in the spectra~\cite{VL2016,PBBCDSY2019}. This is true for both model configurations (weak and strong LFV).

 The BLVs have been computed with the Benettin algorithm~\cite{BGGS1980}, while the CLVs and their adjoint have been computed by the method seeking to find the intersection of the subspaces spanned by the BLVs and the Forward Lyapunov Vectors (FLVs)~\cite{Legras1995}, see~\ref{sec:appLyap}. To this end, the FLVs have also been computed using the Benettin algorithm. 
 
\subsection{The Koopman (KM) and Perron-Frobenius (PF) eigenfunctions}
\label{sec:DMDs}

\subsubsection{Koopman and the Perron-Frobenius operators}
\label{sec:KPFoperator}

The Koopman operator provides a means of representing a finite-dimensional nonlinear system as an infinite-dimensional linear system by `lifting' the underlying state space to a set of observables. The Koopman operator $\mathcal{K}^{\tau}$ acts upon an observable $g(\vec x)$ of the system state $\vec x$ as,
\begin{equation}
    \mathcal{K}^{\tau} \, g(\vec x) = g\left(\vec\Phi^{\tau} (\vec x)\right)
\end{equation}
where the mapping $\vec\Phi^{\tau}$ describes the flow of the system~\refp{eq:sysdyn} such that $\vec x(t+\tau) = \vec\Phi^{\tau} (\vec x(t))$.

While the Koopman operator governs the time evolution of observables of the system, its adjoint, the Perron-Frobenius (or transfer) operator $\mathcal{P}^{\tau}$, governs the time evolution of the probability density $\rho_t$. The probability density given at any lead time $\tau$ is thus,
\begin{equation}
    \label{eq:PFoperator}
    \rho_{t+\tau} = \mathcal{P}^{\tau} \, \rho_t .
\end{equation}
The Koopman and Perron-Frobenius operators can both be used to determine the evolution of the expected value of an observable. Indeed, if we consider the expected value of an observable $g$, for a given distribution $\rho_t$ at time $t$, to be defined as,
\begin{align}
    \langle g \rangle_t = \int g(\vec x) \, \rho_t(\vec x) \, \mathrm{d}\vec x
\end{align}

and the inner product is defined as,
\begin{equation}
    \label{eq:scalarprod}
    \langle a, b \rangle = \int a^\ast (\vec x) \, b(\vec x) \, \mathrm{d}\vec x,
\end{equation} 
then for a real-valued scalar observable $g$, we have $\langle g \rangle_t = \langle g, \rho_t \rangle \nonumber = \langle g, \mathcal{P}^{t} \rho_0 \rangle = \langle \mathcal{K}^{t} g, \rho_0 \rangle$.
Note that we have used the fact that the observable is real, i.e. $g(\vec x)^\ast = g(\vec x)$, and that $\mathcal{K}^t$ is the adjoint of $\mathcal{P}^t$.

For the remainder, for the sake of simplicity, we shall assume that the spectra of the Koopman and Perron-Frobenius operators are discrete.\footnote{However, special care must be taken when the system is chaotic, as it may include degenerate eigenvalues (Jordan blocks) and continuous parts~\cite{GNPT1995, AM2017, M2020}. See also the conclusion (Section~\ref{sec:conclusions}).} Importantly, a vector-valued observable $\vec g$ can then be decomposed using the eigenfunctions $\phi_i$ of the Koopman operator
\begin{equation}
    \vec g(\vec x) = \sum_{i=1}^\infty \, \vec c_i^{\rm KM}  \, \phi_i(\vec x)
\end{equation}

and the application of the Koopman operator can thus be decomposed into a set of eigenvalues $\lambda_i$, eigenfunctions $\phi_i$, and modes (coefficients) $\vec c_i^{\rm KM}$ as,
\begin{equation}
    \label{eq:Koopman_dec}
    \mathcal{K}^{\tau} \vec g(\vec x) = \sum_{i=1}^{\infty} \, \vec c_i^{\rm KM} \, \lambda_i(\tau) \phi_i(\vec x). 
\end{equation}
This indicates that the propagation of an observable due to the Koopman operator can be represented as a superposition of oscillating stretching/contracting factors applied to the Koopman eigenfunctions. A challenge in translating the use of this Koopman operator to practical applications is the determination of an appropriate truncation of this infinite series.

Similarly for the Perron-Frobenius operator, a probability density $\rho$ defined over the phase space can be expanded in terms of its eigenfunctions $\psi_i$:
\begin{equation}
    \label{eq:rhodec}
    \rho(\vec x) = \sum_{i=1}^\infty \, c_i^{\rm PF} \, \psi_i(\vec x) 
\end{equation}
and its time evolution is then also decomposable in term of a set of eigenvalues $\lambda_i$, eigenfunctions $\psi_i$, and coefficients $c_i^{\rm PF}$:
\begin{equation}
    \mathcal{P}^{\tau}  \rho(\vec x) = \sum_{i=1}^{\infty} \, c_i^{\rm PF} \, \lambda_i^\ast(\tau) \psi_i(\vec x). 
\end{equation}

The eigenfunctions of the Koopman and Perron-Frobenius operators are biorthonormal to one another $\langle \phi_i, \psi_j \rangle = \delta_{i,j}$, and therefore, the Koopman modes of a given observable $\vec g$ can be determined using the Perron-Frobenius eigenfunctions: $\vec c_i^{\rm KM} =  \langle \psi_i , \vec g \rangle$, where the inner product is applied component-wise.

The time evolution of the expected value of the observable can then be given more simply as
\begin{align}
    \langle \vec g \rangle_t & = \langle \vec g, \mathcal{P}^t \rho_0 \rangle \nonumber \\ 
    & = \sum_{i=1}^{\infty} \, \langle \vec g,  \lambda_i^\ast(t) \psi_i \, c_i^{\rm PF} \rangle \nonumber \\
    & = \sum_{i=1}^{\infty} \lambda_i^\ast(t) \, c_i^{\rm PF} \, \left(\vec c_i^{\rm KM}\right)^\ast
\end{align}
From now on, to present numerical methods to approximate the Koopman and Perron-Frobenius eigenfunctions decompositions, we will consider a set of realizations $\vec g_t = \vec g(\vec x_t)$ of a vector-valued observable $\vec g$ of dimension $P$ evaluated over the system states $\vec x_t$ that are assumed to satisfy,
\begin{equation}
    \label{eq:Koopdyn}
    \vec g_{t+\tau} = \mathcal{K}^{\tau} \vec g_{t}.
\end{equation}
If this time evolution is repeated sequentially with a fixed lead time $\tau = \Delta t$, then it constitutes thus a dataset of $K+1$ input-output pairs $(\vec g_k, \vec g_{k+1})$, $k=0,1,\ldots,K$ of the operator:
\begin{equation}
      \vec g_{k+1} = \mathbb{K} \vec g_k \quad , \qquad  \mathbb{K} \equiv \mathcal{K}^{\Delta t}.
\end{equation}
For example, if the observables $\vec g_k$ are the states of the system~\refp{eq:sysdyn} depicted on Figure~\ref{fig:timevol}, then $\Delta t=10$ MTU, corresponding roughly to one day.

\subsubsection{The Dynamic Mode Decomposition (DMD) algorithm}
\label{sec:DMDdec}

The DMD algorithm is a data-driven approach that provides a linear decomposition of a given signal of input-output pairs into a set of spatial patterns called dynamic modes that are modulated by a damping or growing oscillating factor. The approach was first developed in the climate community under the name LIM~\cite{P1989,penland1995}, with its corresponding linear normal modes, as an extension of the Principle Oscillation Patterns (POP) technique of~\citeA{H1988} and~\citeA{VBFH1988}. It was later rediscovered in the fluid mechanics community by~\citeA{S2010} as an extension of the proper orthogonal decomposition (POD)~\cite{BHL1993}. In the framework of dynamical systems like~\refp{eq:sysdyn}, DMD has been identified as an algorithm to approximate the Koopman operator $\mathcal{K}^\tau$ ~\cite{RMBSH2009, WKR2015} and the Perron-Frobenius operator $P^\tau$~\cite{KKS2016}.

The DMD algorithm identifies two sets of vectors, the \emph{adjoint} DMD modes and the \emph{exact} DMD modes~\cite{TRLBK2014}. The former are approximately related to the eigenfunctions of the Koopman operator, while the latter are related to the Koopman modes. 
 
For this purpose, the input-output pairs $(\vec g_k, \vec g_{k+1})$ are stacked as the columns of two matrices $\mat X = [\vec g_0 \ldots \vec g_{K-1} ]$ and $\mat Y = [\vec g_1 \ldots \vec g_K ]$. When the time steps are evenly partitioned, this is simply a repeated representation of the dataset, offset by one timestep. It is assumed that a matrix $\mat M^{\rm DMD}$ exists that approximates the operator $\mathcal{K}^\tau$ so that, 
\begin{equation}
    \mat{Y} = \mat M^{\rm DMD} \mat{X},
\end{equation}
and thus,
\begin{equation}
    \label{eq:snapevol}
    \mat M^{\rm DMD} = \mat Y \,\mat X^+,
\end{equation}
where $\mat X^+$ is the pseudoinverse of $\mat X$. 
Alternatively, the matrix $\mat M^{\rm DMD}$ is sometimes written
\begin{equation}
    \label{eq:snapevol2}
    \mat M^{\rm DMD} = \mat A \, \mat G^+,
\end{equation}
where $\mat{A} = \mat{Y} \, \mat{X}^\ast$ and $\mat{G} = \mat{X} \, \mat{X}^\ast$~\cite{KNKWKSN2018}. The matrix $\mat M^{\rm DMD}$ approximates the operator $\mathbb{K}$ in the least-squares sense. The eigenvalues and the right eigenvectors of $\mat M^{\rm DMD}$ are called the DMD eigenvalues and DMD modes of the data.
In practice, the eigendecomposition of $\mat M^{\rm DMD}$ can be performed with the SVD~\cite{TRLBK2014}, or using the Arnoldi algorithm~\cite{RMBSH2009}. The SVD is computed as $\mat X = \mat U \mat \Sigma \mat V^\ast$. A truncated form can be defined to permit a reduced dimension form of $\mat M^{\rm DMD}$. In that case, the equation~\refp{eq:snapevol} can be transformed as,

\begin{equation}
    \label{eq:DMDredop}
    \tilde{\mat{M}}^{\rm DMD} = \mat U^\ast \mat M^{\rm DMD} \mat U = \mat U^\ast \mat Y \mat V \mat \Sigma^{-1}.
\end{equation}

The nonzero eigenvalues $\lambda_i^{\rm DMD}$ of $\mat M^{\rm DMD}$ are the same as those of $\tilde{\mat{M}}^{\rm DMD}$. The right eigenvectors $\tilde{\vec v}_i$ of $\tilde{\mat{M}}^{\rm DMD}$ can be used to recover the corresponding right eigenvectors $\vec v_i = \frac{1}{\lambda_i^{\rm DMD}} \mat Y \mat V \mat \Sigma^{-1} \, \tilde{\vec v_i}$ of $\mat M^{\rm DMD}$. The left eigenvectors $\tilde{\vec{w}}_i$ of $\tilde{\mat{M}}^{\rm DMD}$ can be used to recover the left eigenvectors $\vec{w}_i = \mat U \, \tilde{\vec{w_i}}$ of $\mat M^{\rm DMD}$, satisfying the biorthonormality condition,
\begin{equation}
    \label{eq:biorth}
    \vec{w}^\ast_i \, \vec v_j =  \tilde{\vec{w}}^\ast_i \mat U^\ast \, \frac{1}{\lambda_j^{\rm DMD}} \, \mat Y \mat V \mat \Sigma^{-1} \, \tilde{\vec v_j} = \tilde{\vec{w}}^\ast_i \, \frac{1}{\lambda_j^{\rm DMD}} \tilde{\mat{M}}^{\rm DMD} \, \tilde{\vec v_j} = \tilde{\vec{w}}^\ast_i \, \tilde{\vec v_j} = \delta_{i,j}
\end{equation}
where we have assumed that the left and right eigenvectors of $\mat M^{\rm DMD}$ are scaled in order to form biorthonormal bases. The left eigenvectors $\vec{w}_i$ are called the \emph{adjoint} DMD modes, while the right eigenvectors $\vec{v}_i$ are called the \emph{exact} DMD modes~\cite{TRLBK2014}.

The left eigenvectors $\vec{w}_i$ of $\mat{M}^{\rm DMD}$ can be used to produce approximations of the Koopman (KM) eigenfunctions. As shown by~\cite{WKR2015}, if the eigenfunctions of the Koopman operator are approximated as
\begin{equation}
    \label{eq:dictokoop}
    \phi_i(\vec x) \approx \vec{w}_i^\ast \, \vec{g}(\vec x).
\end{equation}
then any observable $\vec h$ can be decomposed according to
\begin{equation}
    \label{eq:obsdec}
    \vec h(\vec x) = \sum_{i=1}^P \vec c^{\rm DMD}_i \, \vec w_i^\ast\, \vec g(\vec x)
\end{equation}
and applying the operator $\mathbb{K}$, we recover a time-discretized vector-valued approximation of Eq.~\refp{eq:Koopman_dec}:
\begin{align}
    \label{eq:obsdecevol}
    \mathbb{K} \vec h(\vec x) & = \sum_{i=1}^P \vec c^{\rm DMD}_i \, \vec w_i^\ast\,  \mathbb{K} \vec g(\vec x) \nonumber \\
    & \approx \sum_{i=1}^P \vec c^{\rm DMD}_i \, \vec w_i^\ast\,  \mat M^{\rm DMD} \, \vec g(\vec x) \nonumber \\
    & = \sum_{i=1}^P \vec c^{\rm DMD}_i \, \lambda_i^{\rm DMD} \, \vec w_i^\ast\,  \vec g(\vec x)
\end{align}
In particular, if the observable $\vec g$ is the identity ($\vec g(\vec x) = \vec x$), then the decomposition~\refp{eq:obsdec} is analogous to a one-term Taylor expansion of $\vec h$~\cite{WKR2015}.
In the remainder, references to the KM eigenfunctions are made under the assumption that they are approximated using this procedure, and due to Eq.~\refp{eq:dictokoop}, it may refer equivalently to the eigenfunctions $\phi_i$ or the left eigenvectors $\vec w_i$.

Finally, the coefficients $\vec c^{\rm DMD}_i$ are provided by the right eigenvectors $\vec v_i$, i.e. the \emph{DMD modes} approximating the Koopman modes $\vec c^{\rm KM}_i$ (see for instance a trivial example in the Section~\ref{sec:ensproj} below where $\vec c^{\rm DMD}_i = \vec v_i$).

\subsubsection{The Perron-Frobenius mode decomposition}
\label{sec:PFMDdec}

Since the Perron-Frobenius operator is the adjoint of the Koopman operator, it is also possible to obtain a finite dimensional representation of the former with this relation, as shown by~\citeA{KKS2016}. Because we are working in the space of observables, we can access the eigenfunctions of the Perron-Frobenius operator using the adjoint property with the Koopman operator (e.g. using the inner product~\refp{eq:scalarprod}).
The finite dimensional representation of the Perron-Frobenius operator is given by,
\begin{equation}
    \label{eq:PFMDdef}
    \mat M^{\rm PFMD} = \mat A^\trans (\mat G^+)^\trans .
\end{equation}
where again $\mat{A} = \mat{Y} \, \mat{X}^\ast$ and $\mat{G} = \mat{X} \, \mat{X}^\ast$. 

Similarly as for the DMD decomposition and the Koopman operator, a distribution $\rho(\vec x)$ can be decomposed on the left eigenvectors $\vec{\omega}_i$ of $\mat M^{\rm PFMD}$ as
\begin{equation}
    \label{eq:PFMDdec}
    \rho(\vec x) = \sum_{i=1}^P c_i^{\rm PFMD} \, \vec{\omega}_i^\ast \, \vec g(\vec x)
\end{equation}
where the eigenfunctions $\psi_i$ of the Perron-Frobenius operator are thus approximated as,
\begin{equation}
    \psi_i (\vec x) \approx \vec{\omega}_i^\ast \, \vec g(\vec x) .
\end{equation}
The decomposition~\refp{eq:PFMDdec} of the densities is thus a time-discretized approximation of Eq.~\refp{eq:rhodec}, and we call it a Perron-Frobenius mode decomposition (PFMD). In the following, references to the PF eigenfunctions are made under the assumption that they are approximated using this procedure, and may refer equivalently to the eigenfunctions $\psi_i$ or the left eigenvectors $\vec \omega_i$.

Finally, we note that the Perron-Frobenius operator being considered here is defined with respect to the invariant distribution of the system~\cite{KNKWKSN2018}, since the matrix $\mat M^{\rm PFMD}$ is constructed from a long trajectory of the system dynamics.

\subsubsection{Projections of ensemble distributions}
\label{sec:ensproj}
From now on, we assume that the observable $\vec g$ used to obtain the representations $\mat M^{\rm DMD}$ and $\mat M^{\rm PFMD}$ is the identity: $\vec g(\vec x) = \vec x$. For instance, for a dynamical system, the datasets $\mat X$ and $\mat Y$ considered thus consists of observed states of the system.
For a given ensemble of initial condition perturbations $\vec{\delta x}^m_0$ of the state of the system~\refp{eq:sysdyn}, these can be projected onto a subset of the KM left eigenvectors $\vec{w}_i$ or onto a subset of the PF left eigenvectors $\vec{\omega}_i$. In the first case, it decomposes the perturbations - viewed as local observables - onto selected (approximate) eigenfunctions of the Koopman operator of the system. This subset of eigenfunctions allows one to (approximately) reduce the action of the Koopman operator on a given invariant subspace of this operator, which is characterized by the left eigenvectors $\vec{w}_i$ and the `timescales' $\lambda_i^{\rm DMD}$. The propagation of the projected ensemble of initial conditions by system~\refp{eq:sysdyn} is then assumed to be equivalent to the action of the Koopman operator $\mathcal{K}^t \vec{\delta x}^m_0$ restricted on this invariant subspace.

Let's be more precise about these projections: For a given observable $\vec h$ evaluated on a perturbed state $\vec x + \vec{\delta x}$, we have:
\begin{equation}
  \vec h(\vec x + \vec{\delta x}) \approx \vec h(\vec x) + \vec\nabla_{\vec x} \vec h \, \vec{\delta x}
\end{equation}
The second term is a local approximation of the observable $\vec h$ around the unperturbed state $\vec x$, and whose time evolution is well represented by the DMD decomposition. If the observable $\vec h$ is the identity ($\vec h(\vec x) = \vec x$), we have naturally $\vec\nabla_{\vec x} \vec h = \mat I$ where $\mat I$ is the identity matrix, and $  \vec h(\vec x + \vec{\delta x}) =   \vec h(\vec x) + \vec h(\vec{\delta x})$. Therefore, one can decompose the perturbation according to Eq.~\refp{eq:obsdec} to get:
\begin{equation}
    \vec h(\vec{\delta x}) = \mat{C}^{\rm DMD} \, \mat W^\ast \, \vec g(\vec{\delta x}) = \mat{C}^{\rm DMD} \, \mat W^\ast \, \vec{\delta x}
\end{equation}
where $\mat W$ is the column matrix of left eigenvectors $\vec w_i$ of $\mat M^{\rm DMD}$. Since the observable $\vec h$ is now the identity we have - due to the biorthonormality relationship~\refp{eq:biorth} - that the matrix $\mat{C}^{\rm DMD}$ is given by $\mat{C}^{\rm DMD} = \mat V$ where $\mat V$ is the column matrix of right eigenvectors $\vec v_i$ of $\mat M^{\rm DMD}$. According to Eq.~ \refp{eq:dictokoop}, the decomposition above is a decomposition in terms of the (approximated) eigenfunctions $\phi_i(\vec{\delta x}) \approx (\mat W^\ast \, \vec{\delta x})_i$.
Projecting the perturbation $\vec{\delta x}$ onto a subset of KM eigenfunctions is thus equivalent to making the expansion above according to a partial choice $\mat W'^\ast  \, \vec{\delta x}$ of eigenfunctions, where $\mat W'$ is a column matrix composed of a choice of columns from the matrix $\mat W$, i.e. a choice amongst the left eigenvectors of $\mat M^{\rm DMD}$. The projected perturbation is thus developed as:
\begin{equation}
    \label{eq:obsproj}
    \vec{\delta x}' = \mat V' \, \mat W'^\ast \, \vec{\delta x}
\end{equation}
where $\mat V'$ are the right eigenvectors biorthonormal to the left eigenvectors $\mat W'$. Identifying $\mat B = \mat W'$ in Eq.~\refp{eq:gprojector} for the projector $\mat\Pi$, we get
\begin{equation}
    \mat\Pi = \mat W' \, (\mat W'^\ast \, \mat W')^{-1} \, \mat W'^\ast
\end{equation}
and using the fact that $\mat W'^\ast \, \mat V' = \mat I$, where $\mat I$ is the identity matrix, we have $\mat V' = \mat W' \, (\mat W'^\ast \, \mat W')^{-1}$. Therefore, Eq.~\refp{eq:obsproj} is exactly the projected perturbation $\vec{\delta x}' = \mat\Pi \, \vec{\delta x}$ obtained with the projector~\refp{eq:gprojector}.

Similarly, in the case where the perturbations are projected onto a selected subset of the (approximated) eigenfunctions $\psi_i (\vec{\delta x}) \approx \vec{\omega}_i^\ast \, \vec g(\vec{\delta x})$ of the Perron-Frobenius operator of the system, because this subset forms an invariant subspace of the Perron-Frobenius operator, one can assume that the propagation of the projected ensemble of initial conditions with system~\refp{eq:sysdyn} is equivalent to the action of the Perron-Frobenius operator on the projection of the distribution $\rho^{\rm ens}$ of the ensemble.

\subsubsection{The Koopman and Perron-Frobenius eigenfunctions of the coupled ocean-atmosphere model}

\begin{figure*}
\includegraphics[width=\textwidth]{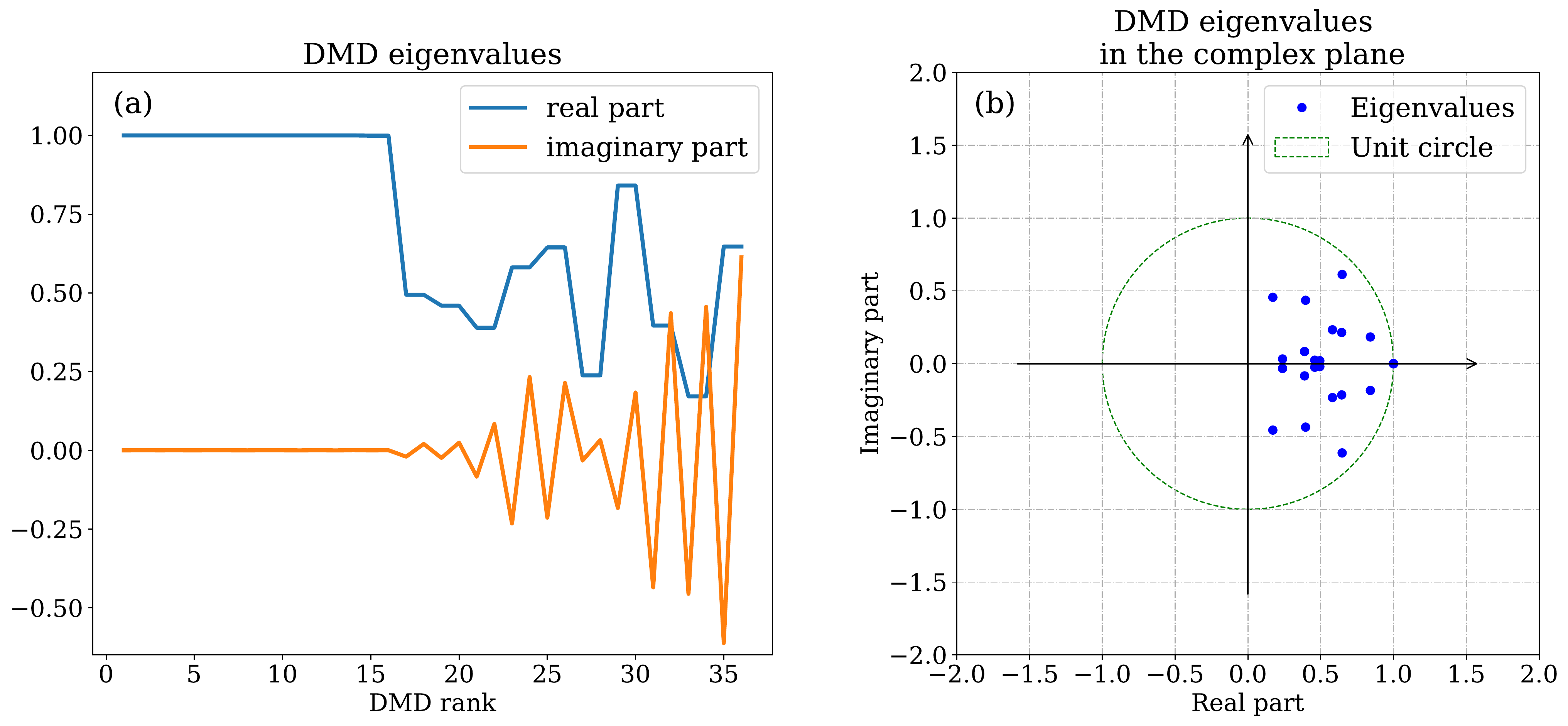}
\caption{Koopman eigenvalues estimated using DMD for the case without low-frequency variability.}
\label{fig:DMDeig_noLFV}
\end{figure*}

\begin{figure*}
\includegraphics[width=\textwidth]{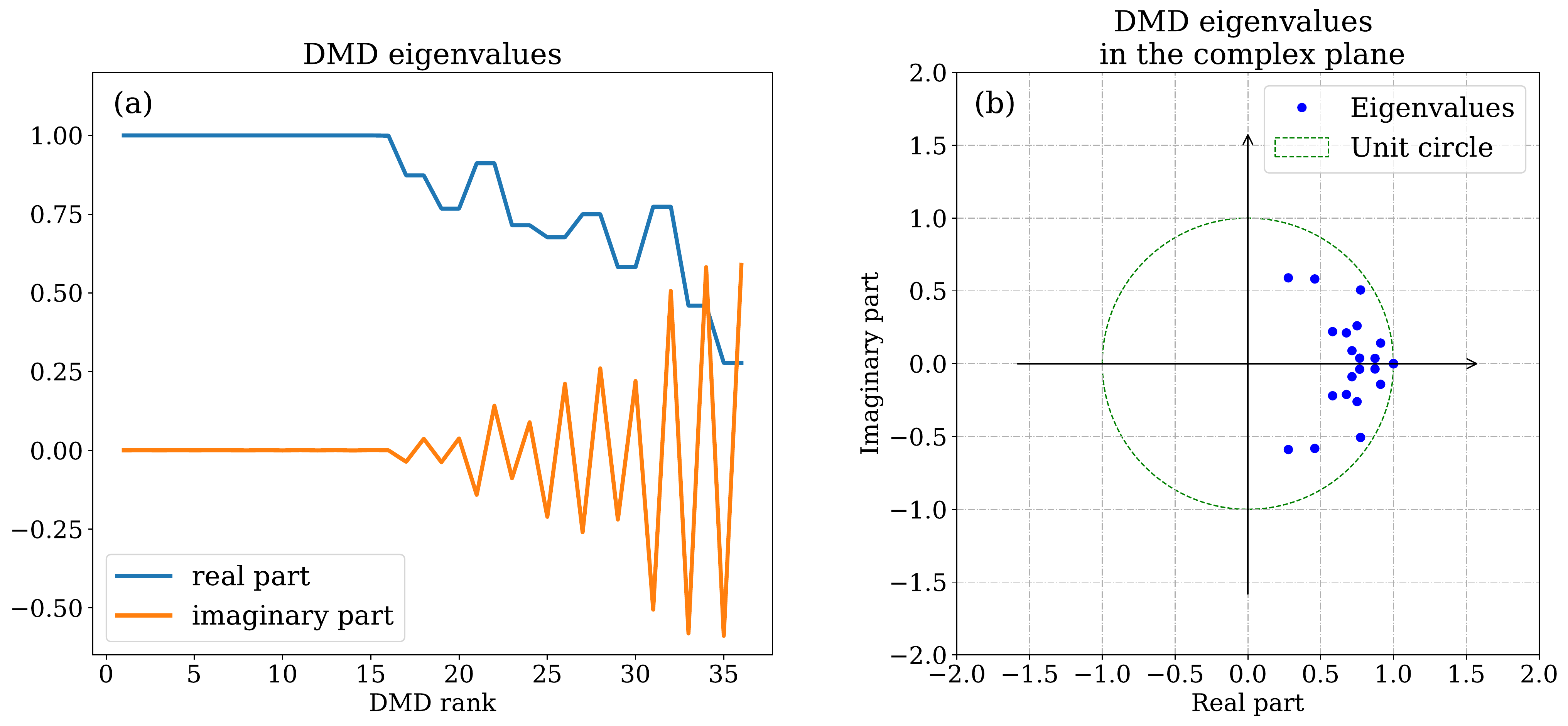}
\caption{Koopman eigenvalues estimated using DMD for the case with low-frequency variability.}
\label{fig:DMDeig}
\end{figure*}

\begin{figure*}
  \includegraphics[width=0.49\textwidth]{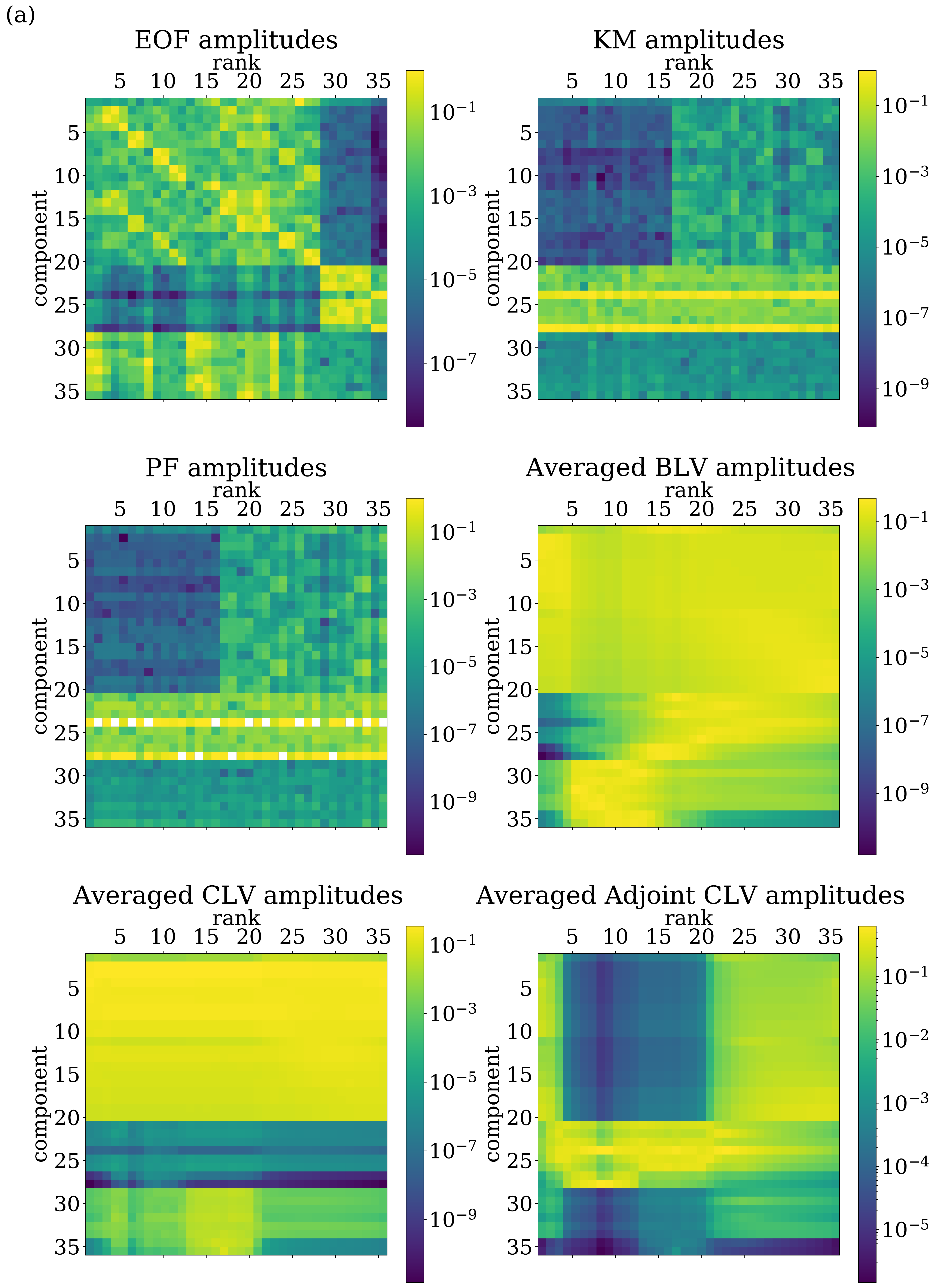}
  \vline \vspace{2pt}
  \includegraphics[width=0.49\textwidth]{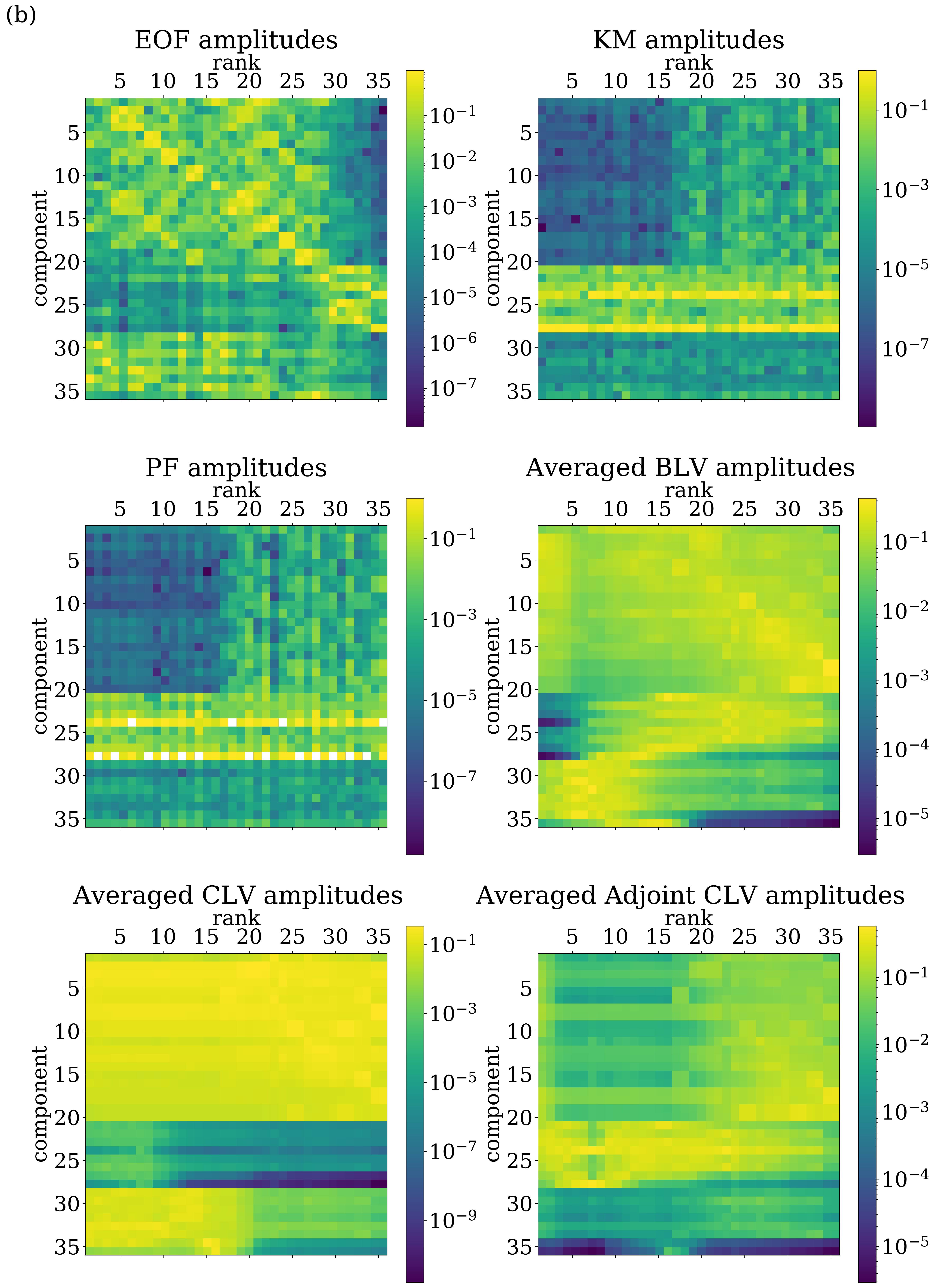}
\caption{Comparison of the averaged energy of the Lyapunov vectors components with the EOF, KM, and PF eigenfunctions patterns for: (a) the case with a weak low-frequency variability, (b) the case with a strong low-frequency variability.}
\label{fig:modes}
\end{figure*}

To study the dynamic modes in the coupled ocean-atmosphere system, the KM eigenfunctions have been estimated using the data of the reference trajectories depicted in Figure~\ref{fig:timevol} sampled every $\Delta t=10$ MTU (roughly every day), using the SVD method described in Section~\ref{sec:DMDdec}. The results are shown in Figure~\ref{fig:DMDeig_noLFV} for the weak LFV case, and in Figure~\ref{fig:DMDeig} for the strong LFV case. In both cases, we note that there are 16 eigenvalues in the vicinity of the point 1 + 0 i in the complex plane. These eigenvalues correspond to very slow decaying and oscillating KM eigenfunctions, describing the LFV signal in the system. The remaining eigenvalues are related to faster decaying oscillations. The amplitude of each component of the KM eigenfunctions is shown in Figure~\ref{fig:modes}.
Each KM eigenfunction is a complex-valued vector, and is paired with another KM eigenfunction that is its complex conjugate (except for the presence of real eigenvalues), each corresponding to complex conjugate eigenvalues. For this reason, Figure~\ref{fig:modes} shows both the real and imaginary parts of the KM eigenfunctions every two columns.
A clear distinction can be made between the slow decaying KM eigenfunctions and the others. Indeed, the slow-decaying KM eigenfunctions (1 to 16) involve both the ocean streamfunction variables (variables 21 to 28) and temperature variables (variables 29 to 36), with a predominance of the streamfunction variables. The fast-decaying KM eigenfunctions (17 to 36) involve the ocean streamfunction variables with a coupling to the atmospheric variables (variables 1 to 20), and a far weaker coupling to the ocean temperature variables.

The PF eigenfunctions have been obtained by directly computing the eigenvectors of the finite dimensional representation of the Perron-Frobenius operator. They possess the same spectrum of eigenvalues as the KM eigenfunctions (see the Supplementary Materials), and while being different, they share the same global slow-fast organization as the KM eigenfunctions (see Figure~\ref{fig:modes}).

\setlength\arrayrulewidth{0.5pt}

\section{Selected bases for experiments}
\label{sec:bases}

Finally, we choose a specific set of bases using the methods described above. Each of the methods determine a set of basis vectors that define the entire state space. We will split each of these into subspaces onto which to project the $M$ `perfect' ensemble perturbations. The projected perturbations hence obtained will be used to obtain the ensemble initial conditions of each experiment in the next section, according to the formula~\refp{eq:pertIC}.

\begin{figure*}
\includegraphics[width=0.495\textwidth]{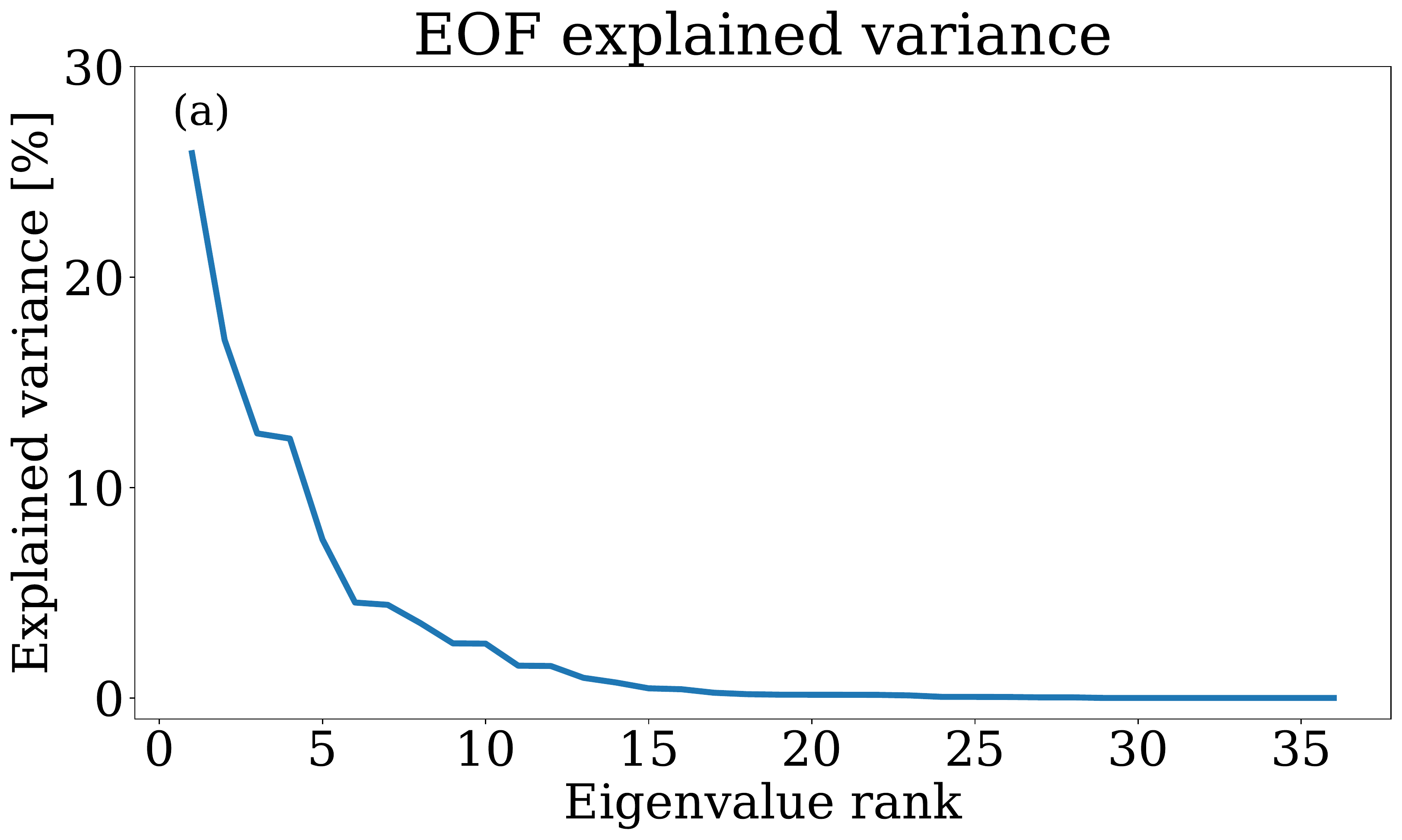}
\includegraphics[width=0.495\textwidth]{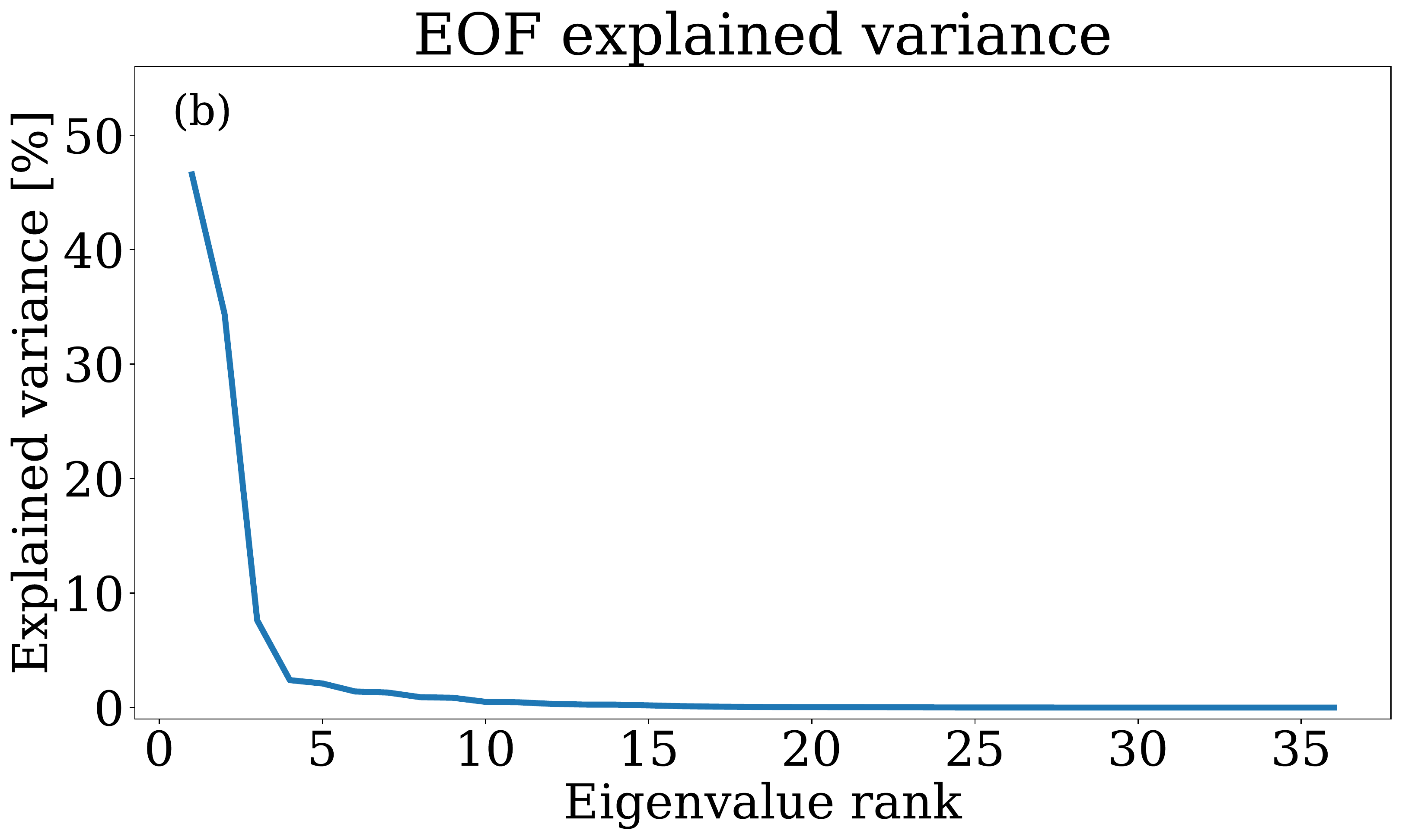}
\caption{Percentage of explained variance of the Empirical Orthogonal Functions modes, for: (a) the case without low-frequency variability and (b) the case with low-frequency variability.}
\label{fig:EOFs}
\end{figure*}

In Figure ~\ref{fig:EOFs}, we show the estimated percent explained variance for the EOFs of the ocean-atmosphere coupled quasi-geostrophic system. 
The EOF modes are shown in Figure~\ref{fig:modes}, with each mode independently normalized to unit magnitude. The leading EOFs explaining most of the variance are related to the ocean temperature and the atmospheric streamfunction variables. We note that the last 8 EOFs, while explaining very little of the total variance, have a qualitatively different pattern than the other modes, with a dominant component along the ocean streamfunction. Therefore, the different bases $\mat B$ of EOFs that we have selected for the experiments are the following:
\begin{itemize}
    \item The first 12 EOFs, $\mat U_{1:12}$, which account for the most significant part of the variability
    \item The last 8 EOFs, $\mat U_{29:36}$, which have a qualitatively different pattern from the others
    \item The remaining 16 EOFs, $\mat U_{13:28}$, which display a more uniform distribution across the different model fields
\end{itemize}

We shall somewhat arbitrarily divide the CLVs, adjoint CLVs, BLVs into 2 parts:
\begin{itemize}
    \item the first $k$ Lyapunov vectors, $\vec\varphi_{1:k}$, $\tilde{\vec{\varphi}}_{1:k}$ and $\vec\varphi^-_{1:k}$, of the spectra, and
    \item the remaining $d-k$ Lyapunov vectors, $\vec\varphi_{(k+1):d}$, $\tilde{\vec{\varphi}}_{(k+1):d}$ and  $\vec\varphi^-_{(k+1):d}$.
\end{itemize}
 The value of $k$ is chosen so that the subspaces $S^-_k$ hence considered includes the unstable directions $\sigma_i > 10^{-2}$ and the near-neutral directions $\sigma_i \in [-10^{-2},10^{-2}]$, see Figure~\ref{fig:Lyapunov}. We shall refer to this as the Unstable Near-Neutral (UNN) subspace. Its complement $H^-_{k+1}$ includes the stable directions $\sigma_i < -10^{-2}$ and will be referred to as the Stable subspace. The subspace $S^-_k$ has been shown to be important for data-assimilation; its dimension $k$ is related to the minimum number of ensemble members needed to ensure that, when applied under ideal conditions, the deterministic Ensemble Kalman Filter (EnKF) is non-divergent~\cite{Bocquet2017, Bocquet2017b, Penny2017, TCVB2020, CBDGRV2021}. This separation of the tangent space into two complementary subspaces is indicated in Figure~\ref{fig:Lyapunov} as a vertical dashed green line at the value $k=20$ for the weak LFV case and $k=16$ for the strong LFV case.

We shall call as \emph{slow} the first 16 KM eigenfunctions and PF eigenfunctions, while the remaining 20 are called the \emph{fast} KM and PF eigenfunctions. We shall use both the slow and the fast KM/PF eigenfunctions as separate bases $\mat B$ in Eq.~ \refp{eq:gprojector} onto which to project the ensemble perturbations. These eigenfunctions are complex valued, but because they are complex conjugate two-by-two, the resulting projector~\refp{eq:gprojector} is a real matrix\footnote{Some KM eigenfunctions may be purely real, but in this case, the corresponding columns and lines of the projector $\mat\Pi$ are also real.}.

Thus, the approximated KM and PF eigenfunctions, derived from DMD, that are used as a basis for the projection of initial ensemble perturbations are,
\begin{itemize}
    \item The `slow' KM and PF left eigenvectors $\vec w_i$ and $\vec \omega_i$, for $i \in \{0,\ldots,16\}$, corresponding to eigenvalues near 1 + 0 i in the complex plane.
    \item The `fast' KM and PF left eigenvectors $\vec w_j$ and $\vec \omega_j$, for $j \in \{17,\ldots,36\}$, corresponding to complex-valued eigenvalues with magnitude notably smaller than 1. 
\end{itemize}

We note that the dimension spanned by these KM and PF left eigenvectors is close to the number of unstable and near-neutral directions found in the system (see next section). It could also be related to the existence of an invariant manifold which forms what was called the ``backbone" of the attractor by~\citeA{VDDG2015}\footnote{See also~\citeA{DV2017} for more details.}, and whence derives the coupled ocean-atmosphere variability in the system. 

\section{Results on ensemble forecast initialization}
\label{sec:results}
As stated in Section~\ref{sec:expdesign}, a set of $N$ states $\vec x_n$ along the reference trajectories are used to generate $N$ separate ensemble forecasts with the initial conditions~\refp{eq:IC} of the perfect ensemble and the initial conditions~\refp{eq:pertIC} of the \emph{projected} ensembles obtained with the various bases described in Section~\ref{sec:bases}. Each experiment uses ensembles composed of $M=20$ members (including the control run). This ensemble size is assumed sufficient based on the dimension of the UNN subspace, as shown by~\citeA{CBDGRV2021} and~\citeA{TCVB2020}.

We compute $N=1980$ sets of ensemble forecasts for the case with weak LFV, and $N=3554$ sets of ensemble forecasts for the case with strong LFV. For strong LFV, to keep the same statistical significance as with the case of weak LFV, we consider a larger number of ensemble forecasts. This is due to the fact that the statistics over these forecasts are performed on two different regions of the phase space. Indeed, in the case of strong LFV, the set of points $\vec x_n$ used to issue forecasts is further divided into two parts that satisfy either $\theta_{{\rm o}, 2} < 0.08$ or $\theta_{{\rm o}, 2} > 0.12$. The same decomposition of the phase space is used by~\citeA{VD2020} and helps to disentangle the distinctly different dynamical behaviours of the two parts, as seen in Figure~\ref{fig:timevol}(c). In the case $\theta_{{\rm o}, 2} < 0.08$, it corresponds to the lower and smoother part of the attractor, while the case $\theta_{{\rm o}, 2} > 0.12$ corresponds to the higher and more chaotic part. The predictability is higher in the lower part, where the atmospheric activity is greatly reduced. On the contrary, the predictability is lower in the higher part of the attractor where, while still coupled to the ocean, the atmosphere is more active~\cite{VDDG2015}. The results of the ensemble forecasts are thus analyzed separately over these two regions of the attractor, with $N_{\rm low} = 1531$ forecasts for the lower part, and $N_{\rm high}=2023$ for the higher one. Finally, we note that in one or two rare occurrences, forecasts had to be dropped from the statistics because the adjoint CLVs did not exist. This corresponds to peculiar points of the attractor know as \emph{tangencies}~\cite{YTKGCR2009, XP2016} where a subset of the CLVs are almost aligned and therefore the biorthonormal relationship with the adjoint CLVs fails. In this case, we preferred to drop completely the corresponding point $\vec x_n$ and its ensemble forecast from the statistics.

\subsection{DSSS skill scores}
\begin{figure*}
\includegraphics[width=\textwidth]{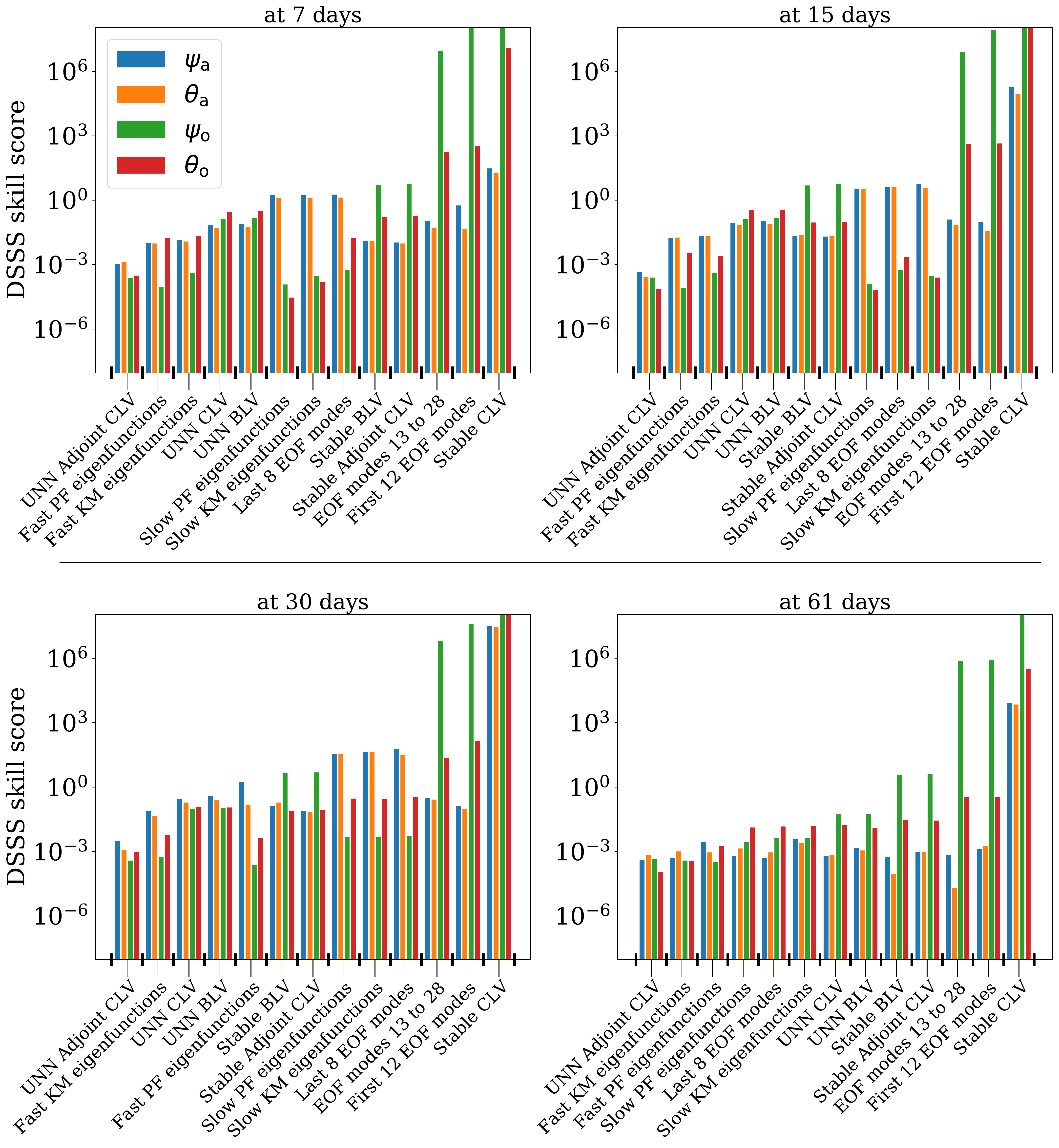}
\caption{DSSS skill score summed over components at different lead times for the case without low-frequency variability. The lower the DSSS score, the better. The methods are sorted by increasing total score value over all four components.}
\label{fig:DSSS_noLFV}
\end{figure*}

\begin{figure*}
\includegraphics[width=\textwidth]{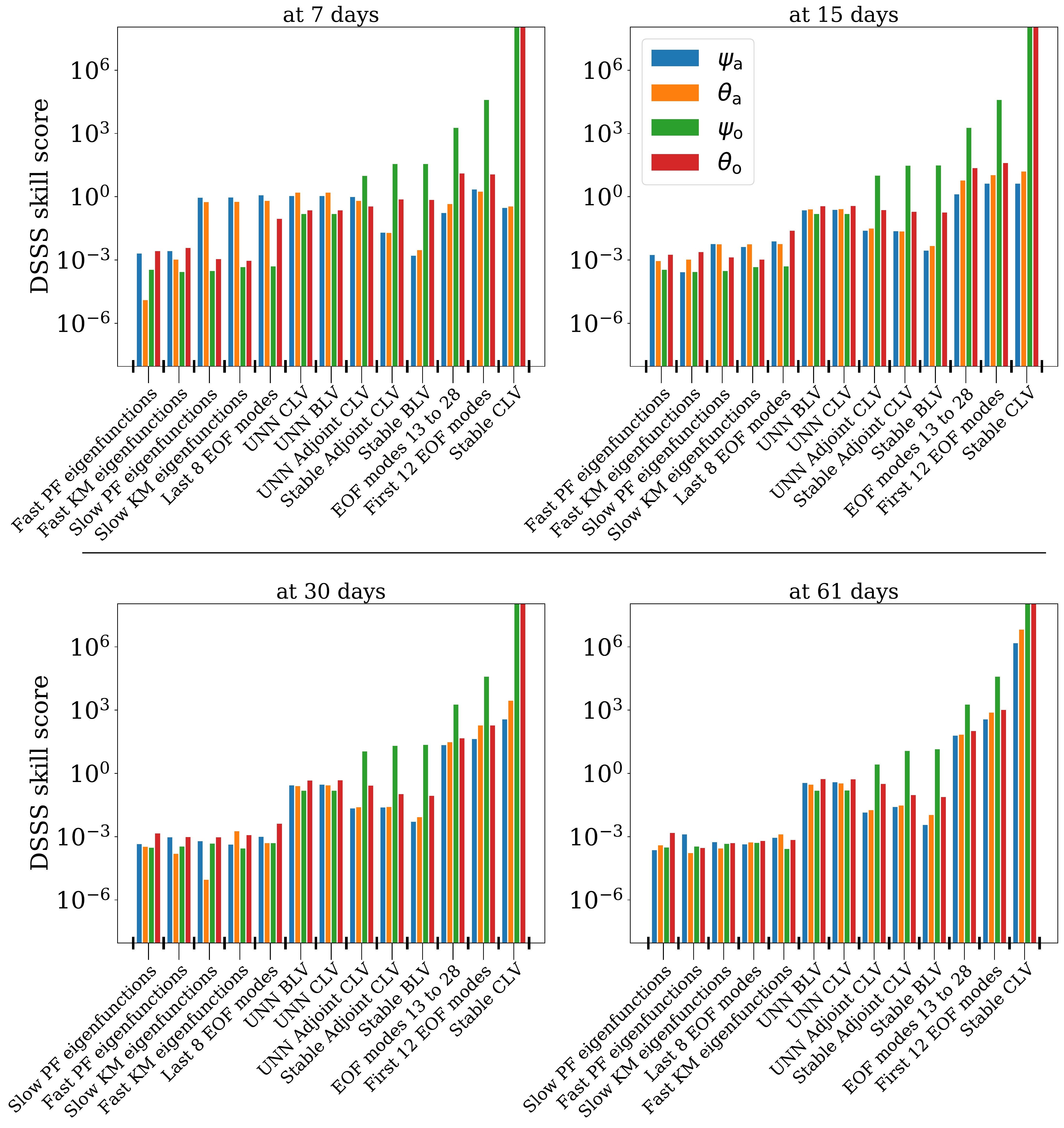}
\caption{DSSS skill score summed over components at different lead times for the case with low-frequency variability and for $\theta_{{\rm o}, 2} < 0.08$. The lower the DSSS score, the better. The methods are sorted by increasing total score value over all four components.}
\label{fig:DSSS_LFV_low}
\end{figure*}

\begin{figure*}
\includegraphics[width=\textwidth]{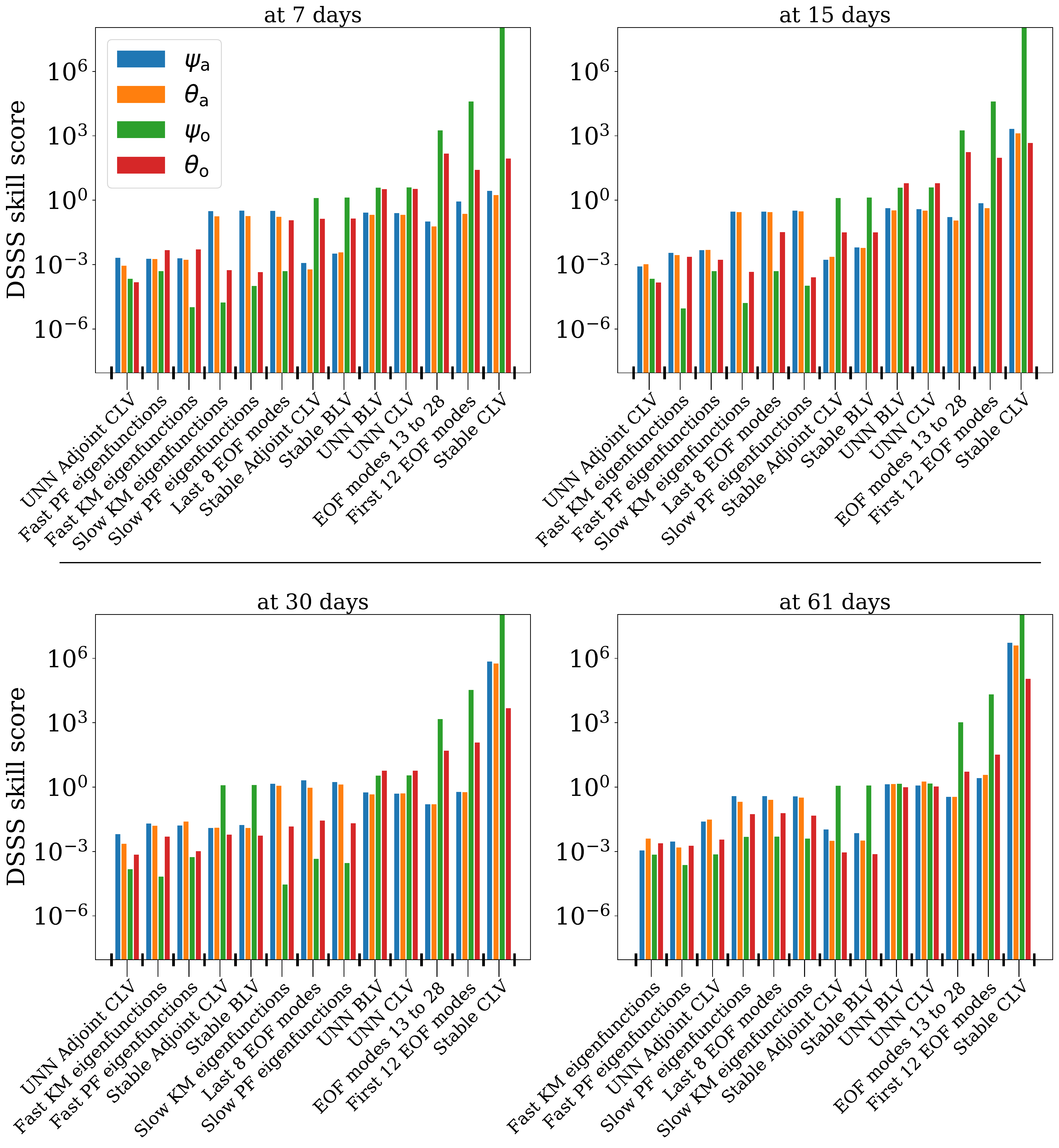}
\caption{DSSS skill score summed over components at different lead times for the case with low-frequency variability and for $\theta_{{\rm o}, 2} > 0.12$. The lower the DSSS score, the better. The methods are sorted by increasing total score value over all four components.}
\label{fig:DSSS_LFV_high}
\end{figure*}

For both experiments with weak and a strong LFV and for all the initial condition projection methods, we show a score based on the DSS discussed in Section~\ref{sec:expdesign}. First we sum the DSS of each variables inside each of the model components to define the components DSS:
\begin{align*}
    \mathrm{DSS}_{\psi_{\rm a}}(\tau) & = \sum_{i=1}^{n_{\rm a}} \, \mathrm{DSS}_{\psi_{{\rm a}, i}}(\tau) \\
    \mathrm{DSS}_{\theta_{\rm a}}(\tau) & = \sum_{i=1}^{n_{\rm a}} \, \mathrm{DSS}_{\theta_{{\rm a}, i}}(\tau) \\
    \mathrm{DSS}_{\psi_{\rm o}}(\tau) & = \sum_{i=1}^{n_{\rm o}} \, \mathrm{DSS}_{\psi_{{\rm o}, i}}(\tau) \\
    \mathrm{DSS}_{\theta_{\rm o}}(\tau) & = \sum_{i=1}^{n_{\rm o}} \, \mathrm{DSS}_{\theta_{{\rm o}, i}}(\tau) .
\end{align*}
For one of the given projection methods detailed in Section~\ref{sec:initmethods}, the Dawid-Sebastiani skill score (DSSS) of each component is then the comparison of the component's DSS of the ensemble forecasts with respect to the component's DSS of the perfect ensemble:
\begin{align*}
    \mathrm{DSSS}^{\rm method}_{\psi_{\rm a}}(\tau) & = 1-\frac{\mathrm{DSS}_{\psi_{\rm a}}^{\rm method}(\tau)}{\mathrm{DSS}_{\psi_{\rm a}}^{\rm perfect}(\tau)} \\
    \mathrm{DSSS}^{\rm method}_{\theta_{\rm a}}(\tau) & = 1-\frac{\mathrm{DSS}_{\theta_{\rm a}}^{\rm method}(\tau)}{\mathrm{DSS}_{\theta_{\rm a}}^{\rm perfect}(\tau)} \\
    \mathrm{DSSS}^{\rm method}_{\psi_{\rm o}}(\tau) & = 1-\frac{\mathrm{DSS}_{\psi_{\rm o}}^{\rm method}(\tau)}{\mathrm{DSS}_{\psi_{\rm o}}^{\rm perfect}(\tau)} \\
    \mathrm{DSSS}^{\rm method}_{\theta_{\rm o}}(\tau) & = 1-\frac{\mathrm{DSS}_{\theta_{\rm o}}^{\rm method}(\tau)}{\mathrm{DSS}_{\theta_{\rm o}}^{\rm perfect}(\tau)} \\
\end{align*}
This skill score is equal to zero if the DSS of the ensemble forecasts obtained with a given projection method have the same DSS as that obtained with the perfect ensemble. The higher the DSSS, the lower the reliability of the forecasts provided by the method. The DSSS skill scores of the ensemble forecasts for the cases with weak LFV, strong LFV on the lower (and less chaotic) part of the attractor, and strong LFV with developed chaos, are shown in Figures~\ref{fig:DSSS_noLFV},~\ref{fig:DSSS_LFV_low}, and~\ref{fig:DSSS_LFV_high}, respectively. In addition, the relation between the ensemble spread and the MSE of the ensemble mean is also provided in the Supplementary Materials. The perturbation methods are sorted using the sum of the DSSS scores of the 4 components of the system from the smallest skill score to the largest, allowing to see at a glance the best methods found. Four lead times are displayed in order to reflect the quality of the methods at both medium-range and sub-seasonal time scales. Three methods consistently show the best performance at these different lead times, namely the use of the UNN adjoint CLVs, the fast-decaying KM and PF eigenfunctions. Better performance is found at medium-range lead times for the UNN adjoint CLVs, while at longer sub-seasonal lead times (61 days) the fast-decaying KM and PF eigenfunctions are better.        

To interpret this feature let us note first that the KM and PF eigenfunctions approximated by DMD are projections of the eigenfunctions of the true Koopman and Perron-Frobenius operators onto the space spanned by the linear monomials (i.e. the full state space), respectively (see sections~\ref{sec:DMDdec} and~\ref{sec:PFMDdec}). It seems therefore that perturbing in the (even approximated) invariant subspaces of these operators of the underlying dynamics is crucial to preserving the statistical properties of the ensemble distributions with respect to the true forecast distributions. 

Similarly, the usefulness of the UNN adjoint CLVs may be related to the fact that the adjoint CLVs can be seen as the eigenfunctions of the Koopman operator defined on the tangent space (see~\ref{sec:appLyap}, Section~\ref{sec:KPFtangent}). Again, projecting the ensemble initial conditions on these adjoint CLVs can be seen as projecting them onto invariant subspaces of the Koopman dynamics on the tangent space. On the contrary, the projections on the CLVs do not provide reliable forecasts. These vectors span the modes of the Koopman operator on the tangent space, that are biorthogonal to the eigenfunctions and are not invariant under the forward action of the operator. This apparently precludes achieving reliable forecasts.

Additional conclusions that can also be drawn from Figures~\ref{fig:DSSS_noLFV},~\ref{fig:DSSS_LFV_low} and~\ref{fig:DSSS_LFV_high} are:  
\begin{itemize}
    \item The EOFs do not provide good overall reliability. In fact, the EOFs generally provide good reliability for only one or two variables, but not for all four simultaneously. For instance, projections of the initial conditions onto the last 8 EOFs provide reliable forecasts for the ocean, onto EOFs 13 to 28 provide reliable forecasts of the ocean streamfunction, and onto the first 12 EOFs provide reliable forecasts for the atmosphere. This behaviour of the forecasts initialized with ensembles projected onto EOFs might be due to the fact that they struggle to represent the coupled nature of the variability of the ocean-atmosphere system.
    \item Projection onto the slow-decaying PF eigenfunctions provides reliable ocean forecasts, but not very reliable atmospheric forecasts. Recall that similarly, the damped normal modes were originally used in early studies with the LIM to predict the evolution of sea surface temperatures in the tropical Pacific. \cite{penland1995}
    \item The fast-decaying PF eigenfunctions provide reliable forecasts, except for the weak LFV experiment at the lead time where the errors saturate (around 30 days). However, they provide the more reliable forecasts in the lower part of the attractor, in the case of a strong LFV.
    \item Projection onto the Unstable and Near-Neutral (UNN) BLVs provide unreliable forecasts, mostly for the ocean temperature, while as shown by~\citeA{VD2020}, the Stable BLVs seem to provide better reliability in the ocean when looking at the relation between the spread and the MSE (see the Supplementary Materials). However, this has to be contrasted with the poor DSSS obtained for these components, which might indicate that the moment of the true forecast distribution is not well represented.
    \item Projection onto the CLVs of both the UNN and Stable subspace provide poor ensemble initial conditions, the former being overdispersive while the latter is underdispersive (see the spread-MSE figures in the Supplementary Materials). In addition to the interpretation given above of the CLVs being similar to Koopman modes defined on the tangent space, we note that these vectors are covariant with the dynamics and therefore might not provide a sufficient dispersion in the directions perpendicular to the flow.
    \item The Stable subspace adjoint CLVs provides reliable forecasts for the atmospheric components, but less reliable oceanic streamfunction forecasts.
\end{itemize}

\subsection{Relationships between the different perturbation subspaces}

The results of the previous section clearly indicate the importance of initializing the ensemble forecast with perturbations that are related with the eigenfunctions of the Koopman and Perron-Frobenius operators. To clarify the usefulness of the different subspaces, the angles between the different types of basis vectors are analyzed.

\begin{figure*}
\includegraphics[width=0.495\textwidth]{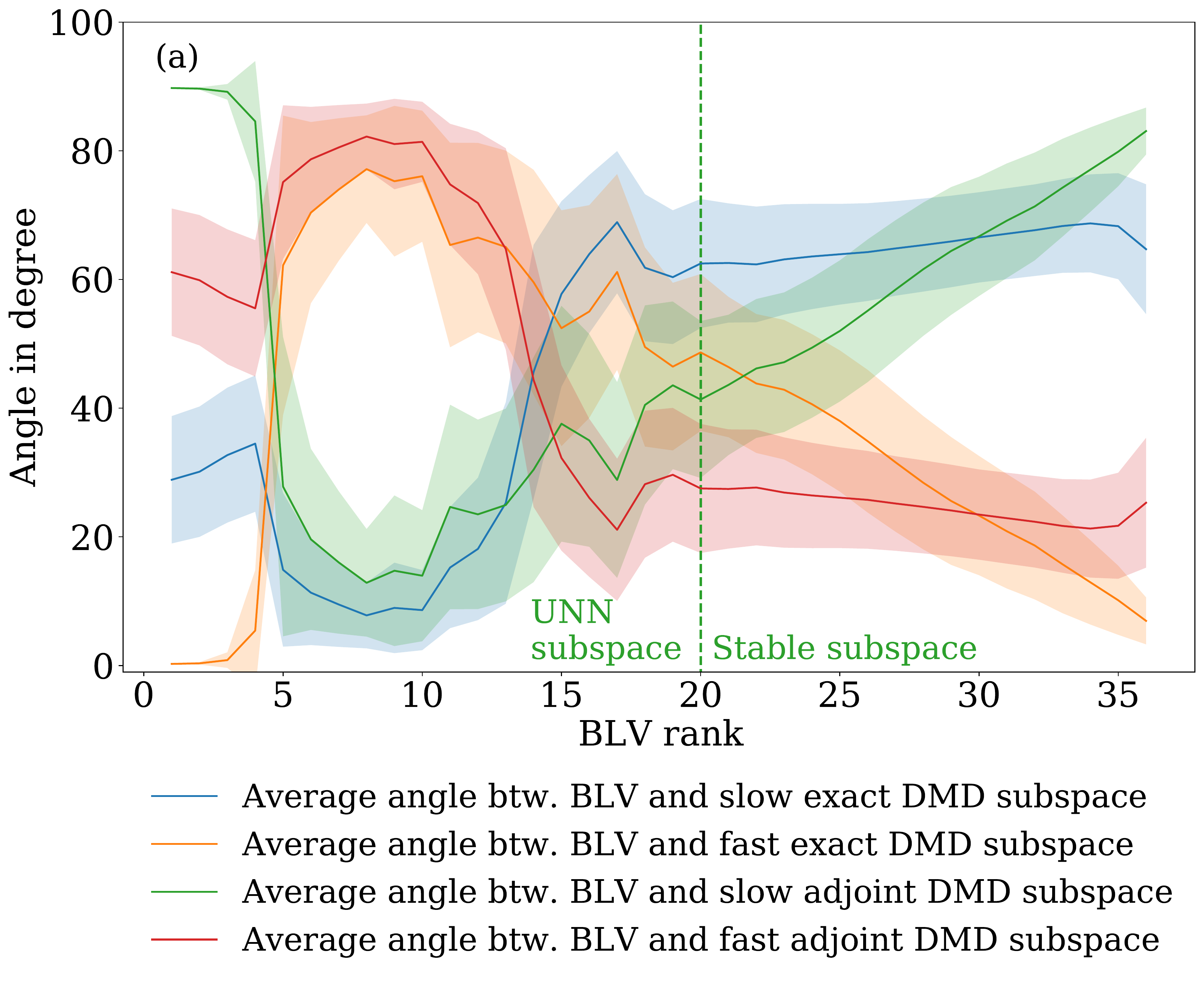}
\includegraphics[width=0.495\textwidth]{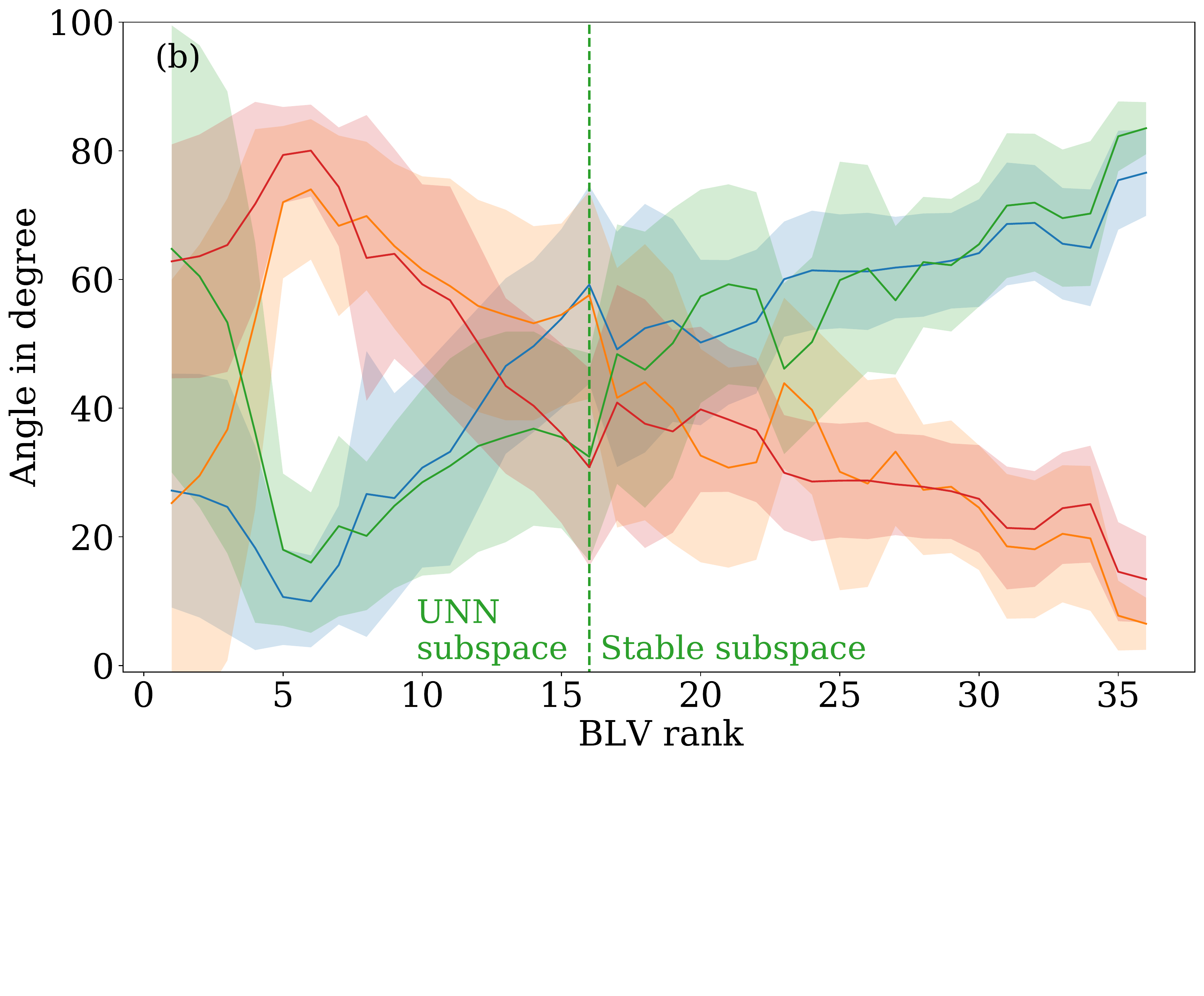}
\caption{Averaged angle in degrees between the Backward Lyapunov Vectors (BLVs) and the Dynamic Modes (DMDs),  for: (a) the case without low-frequency variability and (b) the case with low-frequency variability. The one standard deviation intervals are depicted by the shaded area. The slow and fast exact DMD subspaces are spanned by the right eigenvectors $\vec v_i$ (the DMD modes), for respectively $i \in \{1,\ldots,16\}$ and $i \in \{17,\ldots,36\}$, while the slow and fast adjoint DMD subspaces are spanned by the left eigenvectors $\vec w_i$ (the KM eigenfunctions), again for respectively $i \in \{1,\ldots,16\}$ and $i \in \{17,\ldots,36\}$. See Section~\ref{sec:bases} for an explanation of the slow-fast separation on the modes and eigenfunctions. Note that due to the biorthormality relationship~\refp{eq:biorth} between the vectors $\vec v_i$ and $\vec w_i$, the slow exact DMD subspace is orthogonal to the fast adjoint DMD subspace, while the fast exact DMD subspace is orthogonal to the slow adjoint DMD subspace. The separation between the BLVs belonging to the UNN and stable subspace is depicted by a vertical dashed line.}
\label{fig:BLVvsDMD}
\end{figure*}

In Figure ~\ref{fig:BLVvsDMD}, the average angle between the BLVs and the various exact (the linear approximation of the Koopman modes) and adjoint DMD (the linear approximation of the 
Koopman eigenfunctions) subspaces is shown. An interesting feature is that the BLVs from 15 to 36 are better aligned with the fast adjoint DMD subspace than the set of BLVs from 1 to 14, providing an alternative explanation 
of the good performance of this set of vectors in ensemble forecasting as discussed and illustrated by~\citeA{VD2020}.  
Note however that the angle between these vectors and the fast adjoint DMD subspace is still not negligible (between 20 to 30 degrees), a quite large misalignment with the fast adjoint DMD, that could explain why the stable BLVs are not as effective as the adjoint DMD modes.

\begin{figure*}
\includegraphics[width=0.495\textwidth]{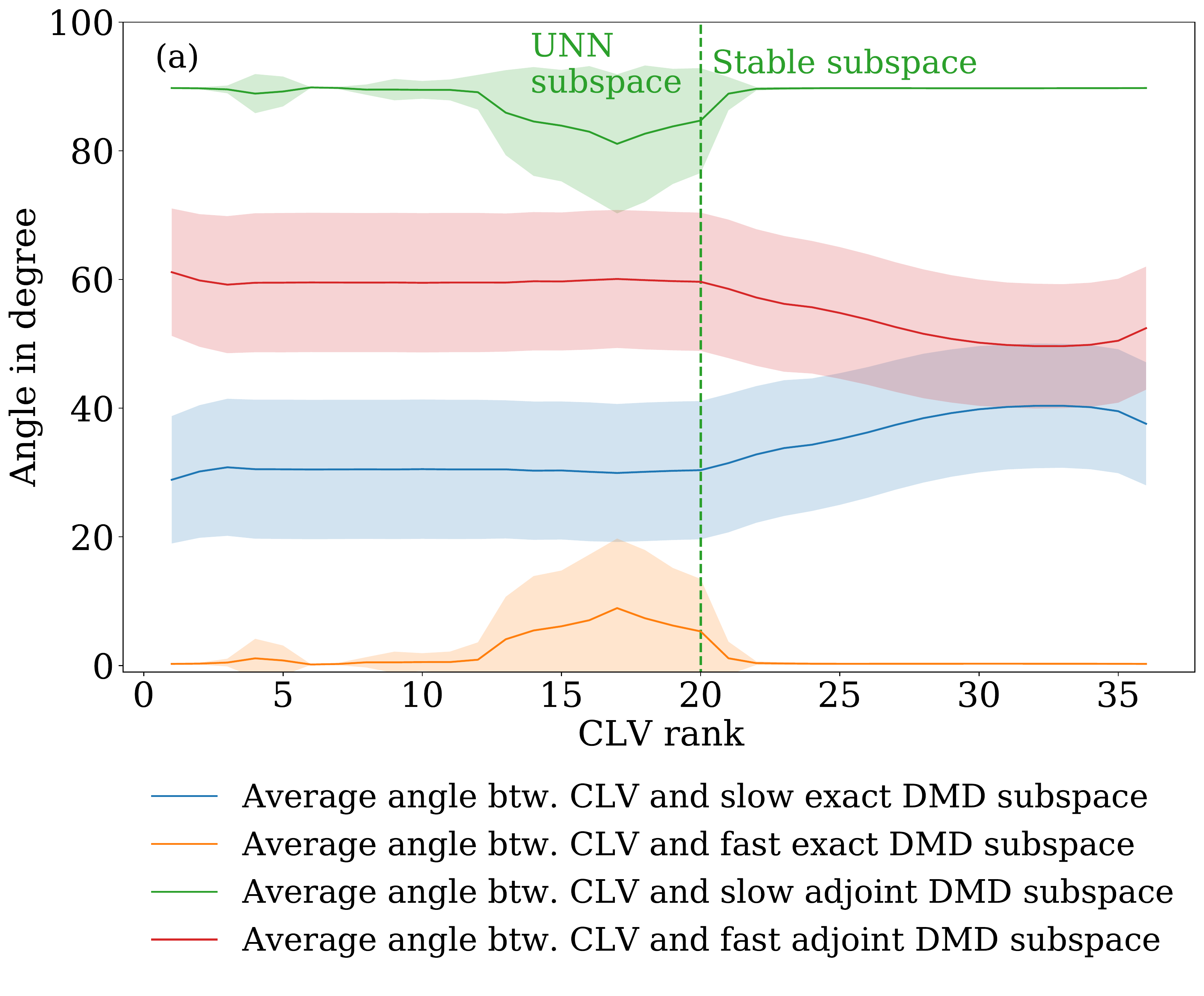}
\includegraphics[width=0.495\textwidth]{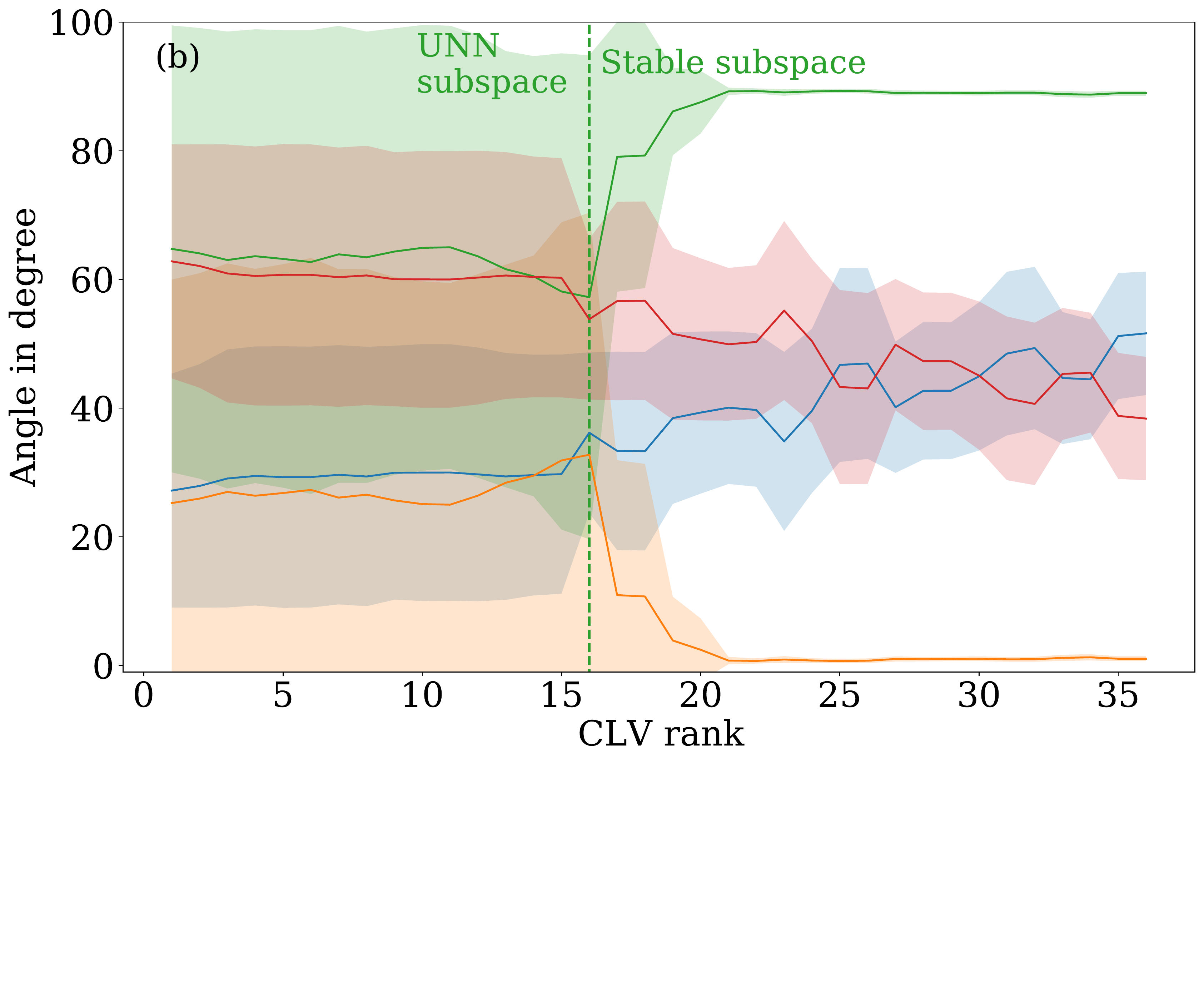}
\caption{Averaged angle in degrees between the Covariant Lyapunov Vectors (CLVs) and the Dynamic Modes (DMDs),  for: (a) the case without low-frequency variability and (b) the case with low-frequency variability. The one standard deviation intervals are depicted by the shaded area. The slow and fast exact DMD subspaces are spanned by the right eigenvectors $\vec v_i$ (the DMD modes), for respectively $i \in \{1,\ldots,16\}$ and $i \in \{17,\ldots,36\}$, while the slow and fast adjoint DMD subspaces are spanned by the left eigenvectors $\vec w_i$ (the KM eigenfunctions), again for respectively $i \in \{1,\ldots,16\}$ and $i \in \{17,\ldots,36\}$. See Section~\ref{sec:bases} for an explanation of the slow-fast separation on the modes and eigenfunctions. Note that due to the biorthormality relationship~\refp{eq:biorth} between the vectors $\vec v_i$ and $\vec w_i$, the slow exact DMD subspace is orthogonal to the fast adjoint DMD subspace, while the fast exact DMD subspace is orthogonal to the slow adjoint DMD subspace. The separation between the CLVs belonging to the UNN and stable subspace is depicted by a vertical dashed line.}
\label{fig:CLVvsDMD}
\end{figure*}

\begin{figure*}
\includegraphics[width=0.495\textwidth]{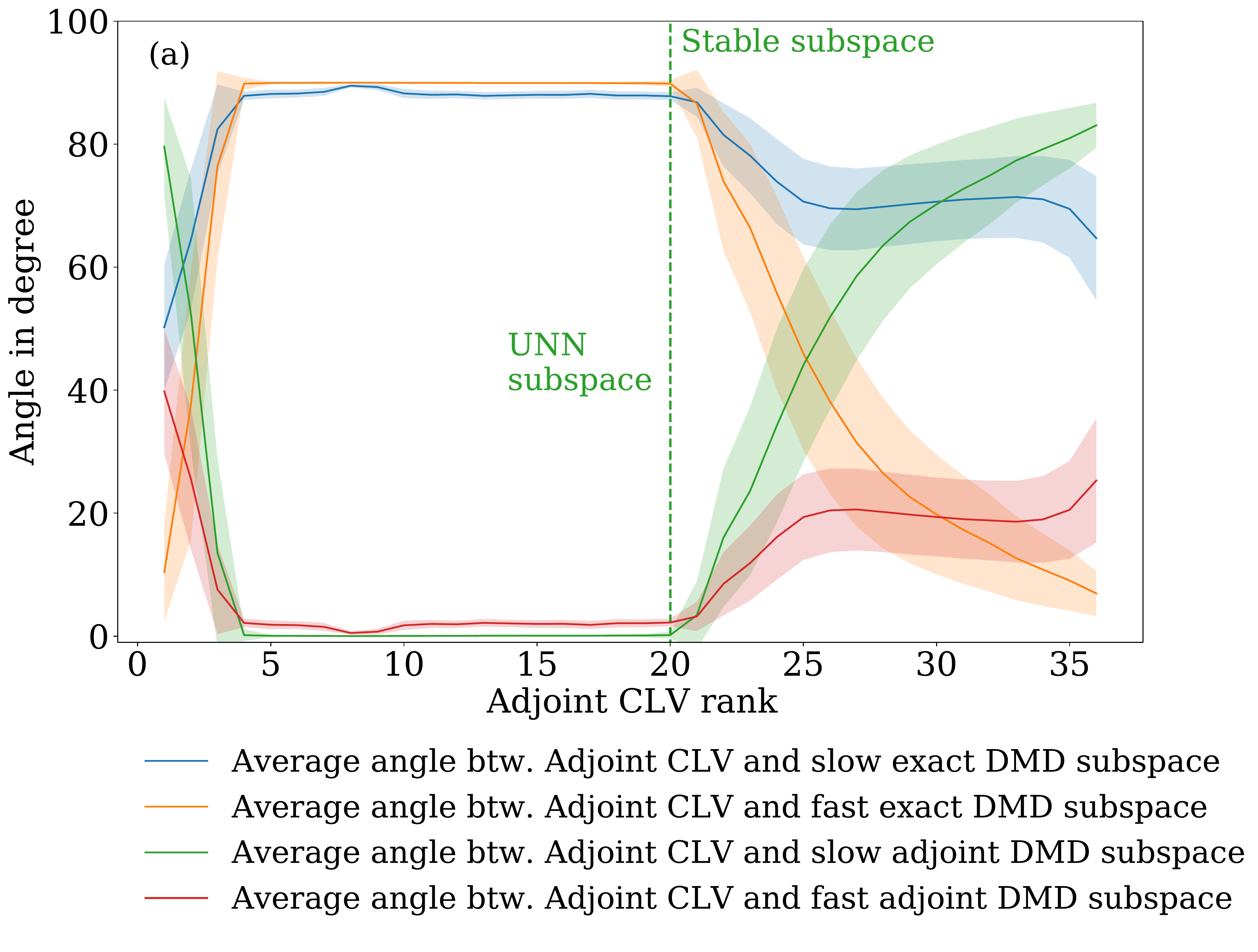}
\includegraphics[width=0.495\textwidth]{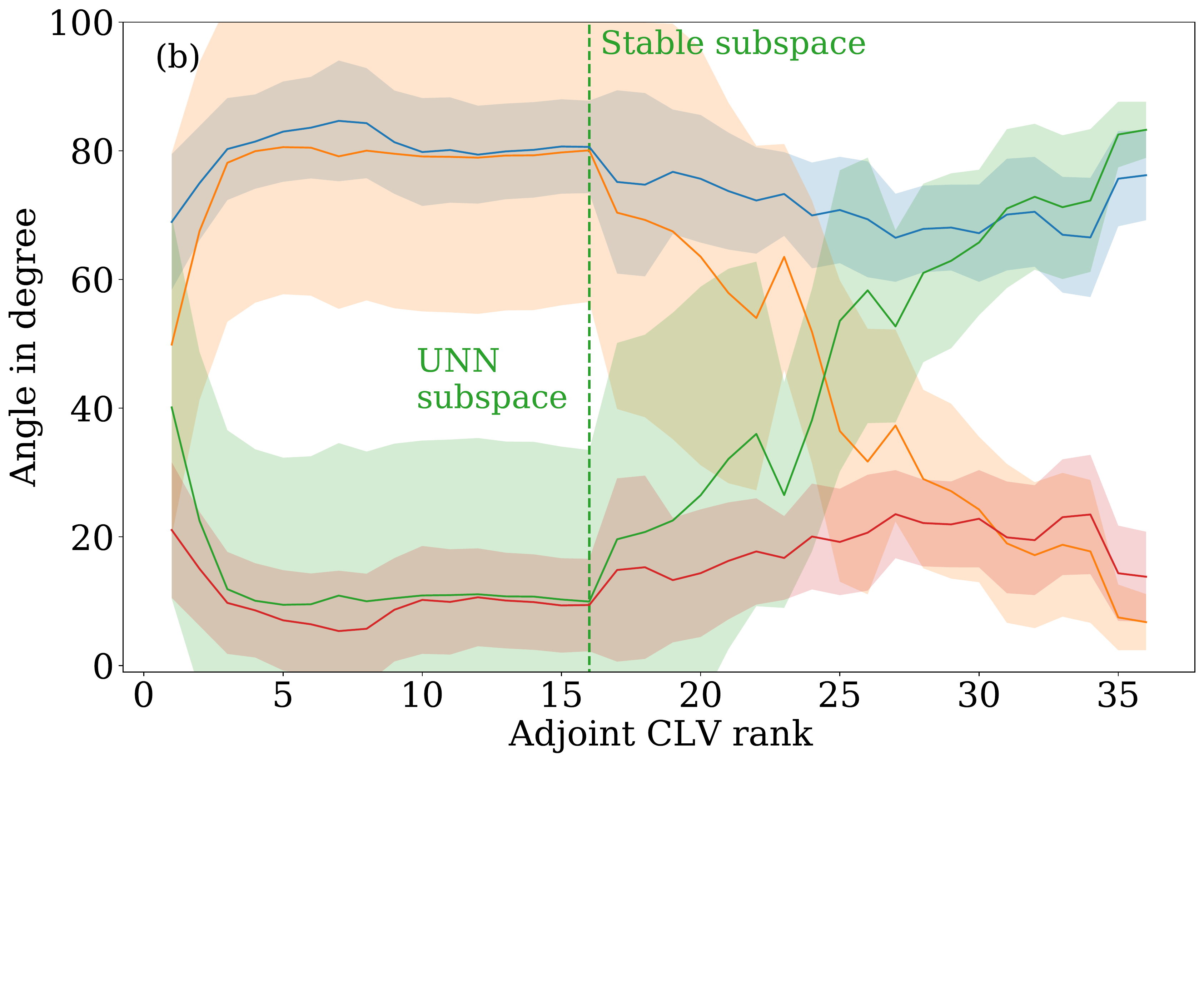}
\caption{Averaged angle in degrees between the adjoint Covariant Lyapunov Vectors (adjoint CLVs) and the Dynamic Modes (DMDs),  for: (a) the case without low-frequency variability and (b) the case with low-frequency variability. The one standard deviation intervals are depicted by the shaded area. The slow and fast exact DMD subspaces are spanned by the right eigenvectors $\vec v_i$ (the DMD modes), for respectively $i \in \{1,\ldots,16\}$ and $i \in \{17,\ldots,36\}$, while the slow and fast adjoint DMD subspaces are spanned by the left eigenvectors $\vec w_i$ (the KM eigenfunctions), again for respectively $i \in \{1,\ldots,16\}$ and $i \in \{17,\ldots,36\}$. See Section~\ref{sec:bases} for an explanation of the slow-fast separation on the modes and eigenfunctions. Note that due to the biorthormality relationship~\refp{eq:biorth} between the vectors $\vec v_i$ and $\vec w_i$, the slow exact DMD subspace is orthogonal to the fast adjoint DMD subspace, while the fast exact DMD subspace is orthogonal to the slow adjoint DMD subspace. The separation between the adjoint CLVs belonging to the UNN and stable subspace is depicted by a vertical dashed line.}
\label{fig:AdCLVvsDMD}
\end{figure*}

An even more interesting result is shown in Figure ~\ref{fig:CLVvsDMD}, in which most of the CLVs are rather well-aligned with the exact DMD subspace (i.e. the space of Koopman modes) and orthogonal to the adjoint DMD subspace (i.e. the space of the Koopman eigenfunctions). This is particularly true for the stable CLVs, which produce unreliable forecasts in our experiments. On the contrary, as shown in Figure~\ref{fig:AdCLVvsDMD}, most of the adjoint CLVs are orthogonal to the exact DMD subspace, and aligned with the adjoint DMD subspace. This is particularly true for the slow UNN adjoint CLVs, which almost entirely align with the adjoint DMD subspace, and provide the most reliable forecasts in our experiments. Similar results have been obtained with the PFMD modes. It thus coherently indicates that the adjoint CLVs are very important structures that can considerably improve the ensemble forecasts. Moreover, a decomposition of the observables in terms of the CLVs and the adjoint CLVs on the tangent linear space yields a similar structure as the one of the DMD decomposition, as shown in~\ref{sec:appLyap}, Section~\ref{sec:KPFtangent}.

Finally, we note that while the UNN adjoint CLVs yield reliable forecasts, the KM and PF eigenfunctions are similar in terms of performance but are much simpler and more straightforward to compute using the DMD algorithm. While the computation of the CLVs typically requires the integration of the tangent linear model over long time periods, both forward and reverse in time, the KM and PF eigenfunctions can be computed from data produced either by numerical simulations, observational analysis products, or reanalysis products, requiring only an efficient algorithm to perform the SVD decomposition.

\section{Conclusions}
\label{sec:conclusions}

In this work, the impact of the choice of the initial perturbations on ensemble forecasts of coupled ocean-atmosphere systems has been addressed by investigating a reduced-order coupled model. Different types of perturbations have been selected, including traditional approaches like the Empirical Orthogonal Functions and the Lyapunov vectors, but also novel approaches based on the Dynamic Mode Decomposition which has been noted in recent years as a reasonable computational approximation of the modes and eigenfunctions of the Koopman and Perron-Frobenius operators~\cite{RMBSH2009,TRLBK2014}. After a detailed analysis of the different definitions of the DMD modes that are found in the literature, their use as a basis for initializing ensemble forecasts was explored.

A key result is that projecting initial perturbations onto the fast-decaying KM eigenfunctions and PF eigenfunctions -- which refer here to linear approximations of the eigenfunctions of the Koopman and Perron-Frobenius operators --  provides reliable ensemble forecasts in the system at hand and at the considered lead times. This further suggests that these {eigenfunctions} are essential for providing reliable ensemble forecasts. Moreover, they seems to be less sensitive to the model's regime and local predictability than other methods. Another important result is the usefulness of the adjoint CLVs, which can be seen as eigenfunctions of the Koopman operator in the tangent space of the system trajectory. 
The adjoint CLVs also provide reliable ensemble forecasts. A key difference between the KM and PK eigenfunctions and the adjoint CLVs lies in the fact that the former are defined globally over the attractor of the system, while the latter are local properties of the flow. In an operational setting, the adjoint CLVs would therefore be quite difficult to compute. On the other hand, it is straightforward to compute an estimate of the KM and PF eigenfunctions directly from data with the DMD method, which provides significant flexibility in their computation and use. 

This thought experiment should now be expanded in a more realistic setting by investigating the use of these tools in intermediate order climate models. 
In this framework, a first research question is related to the validity of the DMD-estimated KM spectrum of the systems being considered: In the present considered system, spectra that are clearly identifiable and separable were found. However, it is known that chaotic systems possess complicated spectra~\cite{AM2017,M2020} which contains continuous components. These complicated spectra might hamper the application of the present method to real datasets or to high-resolution models, the DMD analysis providing too few relevant patterns to work with. This will have to be investigated, notably in systems where the dimension is too high to apply the DMD method directly, and have thus to be reduced first.

Another research question concerns the other sources of uncertainty affecting the ensemble forecasts. As previously noted, systematic errors in the forecast model share roughly equal importance with the specification of initial conditions in producing accurate and reliable forecasts. In an operational setting, it is important to take these systematic model errors into account. A possible path forward is to evaluate the projection of assumed model errors onto the KM or PF eigenfunctions, and randomly perturbing the model in that direction. This question will be explored in the future in the context of the current model.

Finally, a few important steps toward an operational implementation of the DMD approach are still needed: First to investigate the impact of data assimilation on the statistics of the initial error and their projections on the KM and PF eigenfunctions, and second to compare the DMD approach to the singular vector techniques that are often used for ensemble initialization and for the propagation of the error covariances, e.g. \cite{EhrendorferTribbia1997}. These steps are planned in a future investigation.

%
%
%
%
%
%
%
%

\acknowledgments
An earlier version of this manuscript benefited from insightful suggestions from Tom Hamill. J.D. and S.V. acknowledge partial support from ROADMAP, a coordinated JPI-Climate/JPI-Oceans project, financed by the Belgian Science Policy under contract B2/20E/P1/ROADMAP. S.G.P. acknowledges support from the Office of Naval Research (ONR) grants N00014-19-1-2522 and N00014-20-1-2580 and from National Oceanographic and Atmospheric Administration (NOAA) grants NA18NWS4680048, NA19NES4320002, NA20OAR4600277, and NA20NWS4680053. S.G.P. would like to thank Tom Hamill for discussions on S2S prediction, and Cecile Penland for discussions on the LIM. \\

\noindent {\bf Code availability:} The code used to obtain the trajectories and Lyapunov vectors of the \emph{VDDG} model is \textsf{qgs}~\cite{DDV2020}. It is available at \url{https://github.com/Climdyn/qgs} and on Zenodo~\cite{qgs_2021}. A version of the computation notebooks suitable for publication is in preparation and will be released by the end of the review process.

\appendix

\section{Lyapunovs vectors (BLVs, CLVs, and their adjoints)}
\label{sec:appLyap}

In dynamical systems described by a set of ODEs like~\refp{eq:sysdyn}, vectors can be defined to describe the local linear stability around its solutions. These vectors can be obtained by considering the linearization of Eq.~\refp{eq:sysdyn} around such a solution $\vec x(\tau)$:
\begin{equation}
\label{eq:TLM}
    \dot{\vec{\delta x}}(\tau) = \left. \frac{\partial \vec f}{\partial \vec x} \right|_{\vec x(\tau)} \, \vec{\delta x}(\tau)
\end{equation}
where $\partial \vec f / \partial \vec x$ is the Jacobian matrix of $\vec f$. The solution of the linearized equation can be formally written as
\begin{equation}
    \label{eq:TLMsolution}
    \vec{\delta x}(t) = \mat M(t, t_0) \, \vec{\delta x}_0 \qquad , \quad \vec{\delta x}_0 = \vec{\delta x}(t_0)
\end{equation}
where $\mat M(t, t_0) = \vec\nabla_{\vec x(t_0)} \vec\phi^{t-t_0}$ is the fundamental matrix of solutions of the system~\refp{eq:TLM}, i.e. the Jacobian matrix of the flow $\vec\phi^{t-t_0}$ of~\refp{eq:sysdyn}, and is thus identified with the linear propagator that propagates the perturbations in the tangent space of $\vec x(\tau)$ between the times $t_0$ and $t$.

\subsection{Osedelets splitting of the tangent space}
The Osedelets theorem~\cite{O1968,O2008} states that the term $\left(\mat M(t,t_0) \mat M(t, t_0)^\trans \right)^{1/(2(t-t_0))}$ is well defined in the limit $t_0 \to \infty$. Its eigenvectors and the logarithm of its eigenvalues are respectively the Backward Lyapunov Vectors (BLVs) $\vec\varphi^-_i(t)$ at the time $t$ and the Lyapunov exponents $\sigma_i$ of the system. The set of the Lyapunov exponents is sometimes called the Lyapunov spectrum and is assumed here to be sorted in decreasing order. The vectors $\vec\varphi^-_i(t)$ are orthogonal and span a set of subspaces
\begin{align}
    S_j^-(t) & = \mathrm{span} \{ \vec\varphi^-_i(t) | i = 1,2,\ldots,j\} \\
    & \qquad , \quad j=1,\ldots,d \nonumber 
\end{align}
toward which any $j$-arbitrary volume $V_j(t_0)$ ($\dim V_j(t_0) = j$) defined at a time $t_0$ in the far past converges under the action of the propagator\footnote{For the sake of simplicity, we present here the case where there are no degenerate Lyapunov exponents in the spectrum. The general case is presented in~\citeA{KP2012}.}: 
\begin{equation}
    \lim_{t_0\to-\infty} \mat M(t, t_0) V_j(t_0) \subset S^-_j(t).
\end{equation}
By construction, we have $S_1^- \subset S_2^- \subset \ldots \subset S^-_{d-1} \subset S^-_d $ which is called a Osedelets splitting of the tangent space at the time $t$~\cite{KP2012}. The BLVs thus span and describe volumes of the tangent space that are reached asymptotically at a given time by arbitrary volumes defined in the far past, and are thus preserved under the tangent flow
\begin{equation}
    \mat M(t,t_0) S^-_j(t_0) = S^-_j(t) .
\end{equation}

Similarly, one can take the limit of the matrix $\left(\mat M(t,t_0)^\trans \mat M(t, t_0) \right)^{1/(2(t-t_0))}$ for $t\to\infty$ and its eigenvectors are the Forward Lyapunov Vectors (FLVs) $\vec\varphi^+_i(t)$. Its eigenvalues are also the Lyapunov exponents $\sigma_i$. The vectors $\vec\varphi^+_i(t)$ are orthogonal as well and span a set of subspaces
\begin{align}
    S_j^+(t) & = \mathrm{span} \{ \vec\varphi^+_i(t) | i = j,j+1,\ldots,d\} \\
    & \qquad , \quad j=1,\ldots,d \nonumber
\end{align}
toward which any arbitrary $j$-volume $V_j(t)$ ($\dim V_j(t) = j$) defined at a time $t$ in the far future converges under the action of the time-reversed propagator: 
\begin{equation}
    \lim_{t\to\infty} \mat M(t_0, t) V_j(t) \subset S^+_j(t_0).
\end{equation}
By construction, we have the sequence $S_d^+ \subset S^+_{d-1} \subset \ldots \subset S_2^+ \subset S^+_1 $ which forms another Osedelets splitting of the tangent space at the time $t_0$. The FLVs thus span and describe volumes of the tangent space that are reached asymptotically\footnote{Under the evolution of the time-reversed tangent flow.} at a given time by arbitrary volumes defined in the far future. These volumes are thus preserved under the time-reversed tangent flow
\begin{equation}
    \mat M(t_0,t) S^+_j(t) = S^+_j(t_0) .
\end{equation}

\subsection{Covariant Lyapunov Vectors and their adjoint}

The Covariant Lyapunov Vectors (CLVs) are vectors $\vec \varphi_i$ such that when the linear propagator $\mat M$ is applied to them, one obtains
\begin{equation}
    \label{eq:CLVcov}
    \mat M(t,t_0) \, \vec\varphi_i(t_0) = \Lambda_i(t,t_0) \, \vec \varphi_i(t).
\end{equation}
and the linearized dynamics~\refp{eq:TLMsolution} transports the CLVs from a time $t_0$ onto the CLVs at time $t$ further downstream the trajectory $\vec x(\tau)$ by multiplying by a \emph{stretching} factor $\Lambda_i$~\cite{G2005,KP2012}.
The CLVs can thus be shown to be the solutions of the equation
\begin{equation}
    \label{eq:CLVeq}
    \dot{\vec{\varphi}}_i(\tau) = \left. \frac{\partial \vec f}{\partial \vec x} \right|_{\vec x(\tau)} \, \vec \varphi_i(\tau) - \chi_i(\tau) \, \vec \varphi_i(\tau)
\end{equation}
with
\begin{equation}
    \label{eq:stretching}
    \Lambda_i(t,t_0) = \exp\left\{ \int_{t_0}^t \, \chi_i(\tau) \, \mathrm{d}\tau \right\}
\end{equation}
where $\chi_i(\tau)$ is the \emph{local} stretching rate at time $\tau$.
The global Lyapunov exponents of the system are recovered in the limit as $t\to\infty$,
\begin{equation}
    \label{eq:lyapexp}
    \sigma_i = \lim_{t\to\infty} \frac{1}{t} \ln | \Lambda_i(t,t_0) | = \lim_{t\to\infty} \frac{1}{t} \int_{t_0}^t \chi_i(\tau) \, \mathrm{d}\tau .
\end{equation}
By definition, each CLV lies at the intersection between the Osedelets subspaces $S^-_j$ and $S^+_j$~\cite{ER1985},
\begin{equation}
    \vec \varphi_j(t) \in S^+_j(t) \cap S^-_j(t) .
\end{equation}

The linear propagator $\mat M$ can be decomposed in terms of the CLVs $\vec \varphi_i$ and their corresponding stretching factors $\Lambda_i$ as
\begin{equation}
    \label{eq:LyapHomo}
    \mat M(t,t_0) = \sum_{i=1}^d \vec \varphi_i(t) \Lambda_i(t,t_0) \tilde{\vec{\varphi}}_i^\trans(t_0)
\end{equation}
where the vectors $\tilde{\vec{\varphi}}_i$ are the adjoint Covariant Lyapunov Vectors satisfying the biorthonormality relation with the CLVs:
\begin{equation}
    \tilde{\vec{\varphi}}_i^\trans \, \vec \varphi_j = \delta_{i,j}
\end{equation}
at any point of the phase space of the system~\cite{G2005}. 
The adjoint CLVs are solutions of the adjoint of Eq.~\refp{eq:CLVeq},
\begin{equation}
    \dot{\tilde{\vec{\varphi}}}_i(\tau) = \left. \frac{\partial \vec f}{\partial \vec x}^\trans \right|_{\vec x(\tau)} \, \tilde{\vec{\varphi}}_i(\tau) - \chi_i(\tau) \, \tilde{\vec{\varphi}}_i(\tau)
\end{equation}
and are covariant with respect to the adjoint dynamics,
\begin{equation}
    \label{eq:adCLVcov}
    \mat G(t,t_0) \, \tilde{\vec{\varphi}}_i(t_0) = \Lambda_i^{-1}(t,t_0) \, \tilde{\vec{\varphi}}_i(t)
\end{equation}
with $\mat G(t,t_0) = \left(\mat M(t,t_0)^{-1}\right)^\trans$, but they are multiplied by the inverse of the stretching factor~\cite{KP2012}. Note that both Eqs.~\refp{eq:CLVcov} and~\refp{eq:adCLVcov} are time-reversible, with the property imposed by Eq.~\refp{eq:stretching} that the stretching factors are inverted upon time-reversal, $\Lambda_i(t,t_0) = \Lambda_i^{-1}(t_0,t)$.
By definition, each adjoint CLV lies at the intersection between the adjoint Osedelets subspaces $H^+_j$ and $H^-_j$~\cite{ER1985},
\begin{equation}
    \tilde{\vec{\varphi}}_j(t) \in H^+_j(t) \cap H^-_j(t) .
\end{equation}
which are preserved under the adjoint tangent flow.
\begin{align}
    \mat G(t,t_0) H^+_j(t_0) & = H^+_j(t), \\
    \mat G(t,t_0) H^-_j(t_0) & = H^-_j(t).
\end{align}
These subspaces form Osedelets splittings of the tangent space as well,
\begin{align}
    H_j^+(t) & = \mathrm{span} \{ \vec\varphi^+_i(t) | i = 1,2,\ldots,j\} \\
    H_j^-(t) & = \mathrm{span} \{ \vec\varphi^-_i(t) | i = j,j+1,\ldots,d\} \\
    & \forall j=1,\ldots,d \nonumber .
\end{align}

To summarize, the BLVs and FLVs can be interpreted as orthonormal basis of vectors defining volumes covariant with the dynamics, while the CLVs and adjoint CLVs define directions in the tangent space that are covariant with the dynamics. \\

\subsection{Koopman operator of the tangent flow}
\label{sec:KPFtangent}
For a given observable $g$ of a system like~\refp{eq:sysdyn}, the time-evolution starting at time $t_0=0$ of the observables in the neighborhood of a given state $\vec x_0$ can be approximated by
\begin{align}
    \label{eq:TLKoopapprox}
    \mathcal{K}^t g(\vec x_0 + \vec{\delta x}_0) & = g\left(\vec\phi^{t}(\vec x_0 + \vec{\delta x}_0)\right) \nonumber \\
    & \approx g\left(\vec\phi^{t}(\vec x_0) + \left(\vec\nabla_{\vec x_0} \vec\phi^{t}\right) \, \vec{\delta x}_0\right) \nonumber \\
    & \approx g\left(\vec\phi^{t}(\vec x_0)\right) + \left(\vec\nabla_{\vec\phi^{t}(\vec x_0)} g \right)^\trans \, \left(\vec\nabla_{\vec x_0} \vec\phi^{t}\right) \, \vec{\delta x}_0 \nonumber \\
    & = g\left(\vec\phi^{t}(\vec x_0)\right) + \left(\vec\nabla_{\vec\phi^{t}(\vec x_0)} g \right)^\trans \, \mat M(t, 0) \, \vec{\delta x}_0
\end{align}
On the other hand, one can naturally define a Koopman operator $\mathcal{T}^t_{\vec x_0}$ on the tangent linear space of a given trajectory $\vec\phi^t(\vec x_0)$, its expression being
\begin{equation}
    \label{eq:TLKoopop}
    \mathcal{T}^t_{\vec x_0} \, \bar g(\vec{\delta x}) = \bar g(\bar{\vec{\phi}}^t_{\vec x_0} (\vec{\delta x})) = \bar g(\mat M(t,0) \, \vec{\delta x})
\end{equation}
where $\bar{\vec{\phi}}^t_{\vec x_0}$ and $\bar g$ are respectively the flow and an observable defined on the tangent linear space.
The action of the Koopman operator $\mathcal{K}^t$ of the system~\refp{eq:sysdyn} approximated by Eq.~\refp{eq:TLKoopapprox} in a neighborhood of $\vec x_0$ can thus be rewritten
\begin{equation}
    \mathcal{K}^t g(\vec x_0 + \vec{\delta x}_0) \approx \mathcal{K}^t g(\vec x_0) +  \mathcal{T}^t_{\vec x_0} \,  \bar g(\vec{\delta x}_0)
\end{equation}
with $\bar g (\vec{\delta x}) = \left(\vec\nabla_{\vec\phi^{t}(\vec x_0)} g \right)^\trans \, \vec{\delta x}$. Using the decomposition~\refp{eq:LyapHomo}, we get then:
\begin{equation}
    \label{eq:Kooplyapdec}
     \mathcal{T}^t_{\vec x_0} \,  \bar g(\vec{\delta x}_0) =  \sum_{i=1}^d \left(\vec\nabla_{\vec\phi^{t}(\vec x_0)} g \right)^\trans \, \vec \varphi_i(t) \Lambda_i(t,0) \tilde{\vec{\varphi}}_i^\trans(0) \, \vec{\delta x}_0
\end{equation}
and from this equation, one can see that similarly to the DMD left eigenvectors for the Koopman operator $\mathcal{K}^t$ presented in section~\ref{sec:DMDdec}, the adjoint CLVs provide an analogy\footnote{In particular, compare Eq.~\refp{eq:TLMKoopeig} with Eq.~\refp{eq:dictokoop}.} for the ``eigenfunctions" of the first-order Koopman operator $\delta{\mathcal{K}}^{t}$, whose representation is provided by the linear propagator $\mat M$. Indeed, if one considers the functions
\begin{equation}
    \label{eq:TLMKoopeig}
    \phi_i^{\rm TL}(\vec{\delta x }, t) = \tilde{\vec{\varphi}}_i^\trans(t) \, \vec{\delta x},
\end{equation}
it is straightforward, using Eq.~\refp{eq:LyapHomo}, that
\begin{align}
    \label{eq:TLMKoopeigevol}
    \mathcal{T}^s_{\vec x_0} \,  \phi_i^{\rm TL}(\vec{\delta x}, t) & = \tilde{\vec{\varphi}}_i^\trans(t) \, \mat M(t,s) \, \vec{\delta x} \nonumber \\
    & = \Lambda_i(t,s) \, \tilde{\vec{\varphi}}_i^\trans(s) \, \vec{\delta x} \nonumber \\
    & = \Lambda_i(t,s) \, \phi_i^{\rm TL}(\vec{\delta x}, s)
\end{align}
On the other hand, the CLVs span the space of the Koopman modes of the operator $\delta{\mathcal{K}}^t$, and one can rewrite Eq.~\refp{eq:Kooplyapdec} as
\begin{equation}
    \label{eq:TLMdec}
    \mathcal{T}^t_{\vec x_0} \,  \bar g(\vec{\delta x}_0) = \sum_{i=1}^d c^{\rm TL}_i(t) \, \Lambda_i(t,0) \, \phi_i^{\rm TL}(\vec{\delta x}_0, 0)
\end{equation}
which is analogous to Eqs.~\refp{eq:Koopman_dec} and~\refp{eq:obsdecevol}. However, note that since the time-evolution in the tangent space is given by a non-autonomous system~\refp{eq:TLM}, both the functions~\refp{eq:TLMKoopeig} and modes $c_i^{\rm TL}(t)=\left(\vec\nabla_{\vec x(t)} g \right)^\trans \, \vec \varphi_i(t)$ of this decomposition are time-dependent.

Finally, due to the similarity between Eq.~\refp{eq:TLMdec} and Eq.~\refp{eq:obsdecevol}, the discussion in Section~\ref{sec:ensproj} about ensemble projections remains appropriate here. Projecting an ensemble of initial conditions $\vec{\delta x}_0^m$ on subspaces spanned by the adjoint CLVs and propagating them is tantamount to projecting on invariant subspaces of the Koopman operator $\mathcal{T}^t_{\vec x_0}$.


%
%

\bibliography{DMDs}

%
%
%
%
%

\newpage
\section*{Supplementary Material : Additional Figures with the MSE, Spread and DSSS as a function of the forecast lead time, and the PFMD spectrum}

\subsection*{Introduction}

\paragraph*{MSE, Spread and DSSS as a function of the forecast lead time:}
In this supplementary note, we show some figures depicting the time-evolution of the scores as a function of the forecast lead time. To recall first the definition of these scores, let's consider a dynamical system
\begin{equation}
  \label{eq:timevol}
  \dot{\vec{x}} = \vec f(\vec x)
\end{equation}
and a set of $n$ ensemble forecasts $\vec y_{m,n}(t)$, $m=1,\ldots,M$ performed with it, $M$ being the size of the ensembles. If $\vec x_n(t)$ is the ``truth'' corresponding to the $n^{\rm th}$ forecast, then the MSE and the Spread of the forecasts are defined as
\begin{align*}
    \mathrm{MSE}(\tau) & = \frac{1}{N} \sum_{n=1}^N \, \left\| \vec{x}_n(\tau) - \bar{\vec{y}}_n(\tau)\right\|^2   \\
    \mathrm{Spread}^2(\tau) & = \frac{1}{N} \sum_{n=1}^N \frac{1}{M-1} \sum_{m=1}^M \, \left\| \vec{y}_{m,n}(\tau) - \bar{\vec{y}}_n(\tau)\right\|^2
\end{align*}
where
\begin{equation*}
    \label{eq:EM}
    \bar{\vec{y}}_n(\tau) = \frac{1}{M} \sum_{m=1}^M \, \vec{y}_{m,n}(\tau) 
\end{equation*}
is the ensemble mean over the members $\vec{y}_{m,n}(\tau)$ of the $n^{\mathrm{th}}$ ensemble forecast. f the $\mathrm{Spread}^2$ and the $\mathrm{MSE}$ are close to one another, indicating that the estimated error is close to the true error, then the ensemble forecast is considered reliable~\cite{LP2008}.

The bias-free univariate DSS for the $n^{\mathrm{th}}$ ensemble forecast and the $i^{\mathrm{th}}$ variable of the system can be written as~\cite{SFSL2019}:
\begin{align*}
    \mathrm{DSS}_{n,i}(\tau) = & \frac{1}{2} \, \log(2\pi) + \frac{1}{2} \, \log \, \sigma_{n,i}^2(\tau) \nonumber \\
    & + \left.\frac{1}{2} \frac{M-3}{M-1} \, \left( \bar{y}_{n,i}(\tau) - x_{n,i}(\tau)\right)^2 \right/ \sigma_{n,i}^2(\tau), \label{eq:DSS}
\end{align*}
where $\sigma_{n,i}^2$ is an estimator of the $i^{\mathrm{th}}$ variable ensemble variance:
\begin{equation*}
    \sigma_{n,i}^2(\tau) = \frac{1}{M-1} \sum_{m=1}^M \, | y_{m,n,i}(\tau) - \bar{y}_{n,i}(\tau)|^2.
\end{equation*}
This score can then be averaged over the $N$ realizations:
\begin{equation*}
    \mathrm{DSS}_i(\tau) = \frac{1}{N} \sum_{n=1}^N \, \mathrm{DSS}_{n,i}(\tau).
\end{equation*}
The lower the DSS score, the more reliable the ensemble forecasts are for this particular variable.

In the context of the MAOOAM-VDDG ocean-atmosphere model considered in the paper, the Dawid-Sebastiani Score (DSS) can be aggregated per component of the system:
\begin{align*}
    \mathrm{DSS}_{\psi_{\rm a}}(\tau) & = \sum_{i=1}^{n_{\rm a}} \, \mathrm{DSS}_{\psi_{{\rm a}, i}}(\tau) \\
    \mathrm{DSS}_{\theta_{\rm a}}(\tau) & = \sum_{i=1}^{n_{\rm a}} \, \mathrm{DSS}_{\theta_{{\rm a}, i}}(\tau) \\
    \mathrm{DSS}_{\psi_{\rm o}}(\tau) & = \sum_{i=1}^{n_{\rm o}} \, \mathrm{DSS}_{\psi_{{\rm o}, i}}(\tau) \\
    \mathrm{DSS}_{\theta_{\rm o}}(\tau) & = \sum_{i=1}^{n_{\rm o}} \, \mathrm{DSS}_{\theta_{{\rm o}, i}}(\tau) .
\end{align*}
where $\psi_{\rm a}$ and $\theta_{\rm a}$ are respectively the streamfunction and temperature of the atmosphere, while $\psi_{\rm o}$ and $\theta_{\rm o}$ are respectively the streamfunction and temperature of the ocean.

Finally, considering several methods to obtain the ensemble forecasts, these aggregated score can be compared to \emph{perfect} ensemble forecasts with the Dawid-Sebastiani Skill Score (DSSS) that we defined as:
\begin{align*}
    \mathrm{DSSS}^{\rm method}_{\psi_{\rm a}}(\tau) & = 1-\frac{\mathrm{DSS}_{\psi_{\rm a}}^{\rm method}(\tau)}{\mathrm{DSS}_{\psi_{\rm a}}^{\rm perfect}(\tau)} \\
    \mathrm{DSSS}^{\rm method}_{\theta_{\rm a}}(\tau) & = 1-\frac{\mathrm{DSS}_{\theta_{\rm a}}^{\rm method}(\tau)}{\mathrm{DSS}_{\theta_{\rm a}}^{\rm perfect}(\tau)} \\
    \mathrm{DSSS}^{\rm method}_{\psi_{\rm o}}(\tau) & = 1-\frac{\mathrm{DSS}_{\psi_{\rm o}}^{\rm method}(\tau)}{\mathrm{DSS}_{\psi_{\rm o}}^{\rm perfect}(\tau)} \\
    \mathrm{DSSS}^{\rm method}_{\theta_{\rm o}}(\tau) & = 1-\frac{\mathrm{DSS}_{\theta_{\rm o}}^{\rm method}(\tau)}{\mathrm{DSS}_{\theta_{\rm o}}^{\rm perfect}(\tau)} \\
\end{align*}
The smaller the DSSS, the better. A value of zero indicates that the considered method matches the perfect ensemble reliability. On the other, a negative value of the DSSS would indicate that the method outperforms the perfect one.

We consider in this supplementary the two different model configurations mentioned in the paper, i.e. one with a weak low-frequency variability (LFV), and one with a strong LFV. In the latter case, we distinguish between two different regions of the attractor: a chaotic region for $\theta_{{\rm o}, 2} >0.12$  and a more "quiet" region for $\theta_{{\rm o}, 2} <0.08$.

\paragraph*{PFMD spectra:}
We also plot the PFMD\footnote{PFMD for Perron-Frobenius Modes Decomposition.} spectra, to show that they are the same as the one obtained with DMD and depicted in the paper.

More precisely, considering two collections of states of the dynamical system~\refp{eq:timevol} $\mat X = [\vec x_0 \ldots \vec x_{K-1}]$ and $\mat Y = [\vec x_1 \ldots \vec x_{K}]$, the PFMD representation of the Perron-Frobenius operator is given by 
\begin{equation}
    \label{eq:PFMDdef}
    \mat M^{\rm PFMD} = \mat A^\trans (\mat G^+)^\trans .
\end{equation}
where $\mat{A} = \mat{Y} \, \mat{X}^\ast$ and $\mat{G} = \mat{X} \, \mat{X}^\ast$. The eigenvalues of the matrix $\mat M^{\rm PFMD}$ form then the above-mentioned spectrum.

\newpage

\subsection*{Experiment the weak LFV}

\paragraph*{MSE and Spread as a function of the forecast lead time}

\begin{center}
   \includegraphics[width=\linewidth]{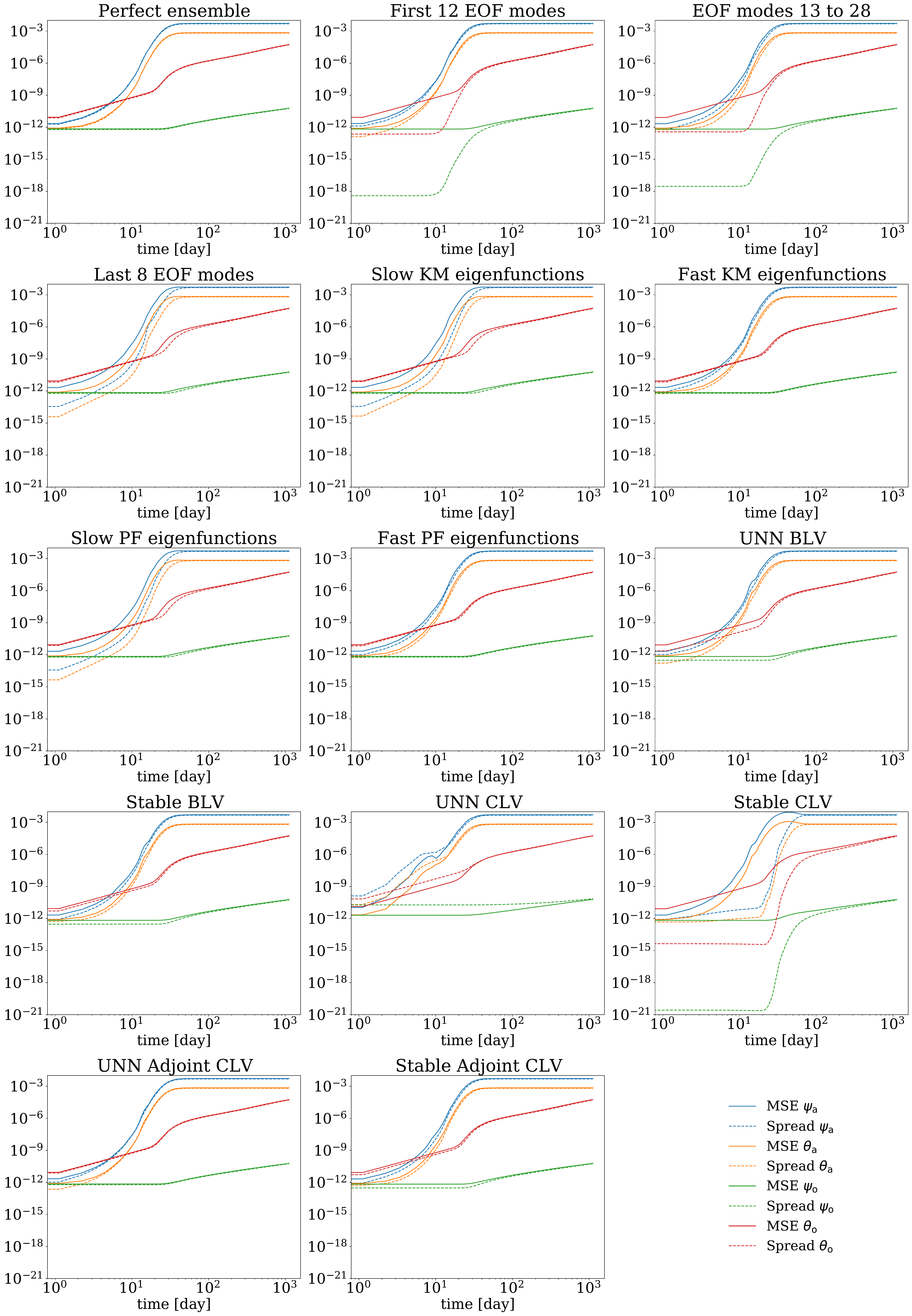}
\end{center}

\newpage

\paragraph*{DSSS as a function of the forecast lead time}

\begin{center}
   \includegraphics[width=\linewidth]{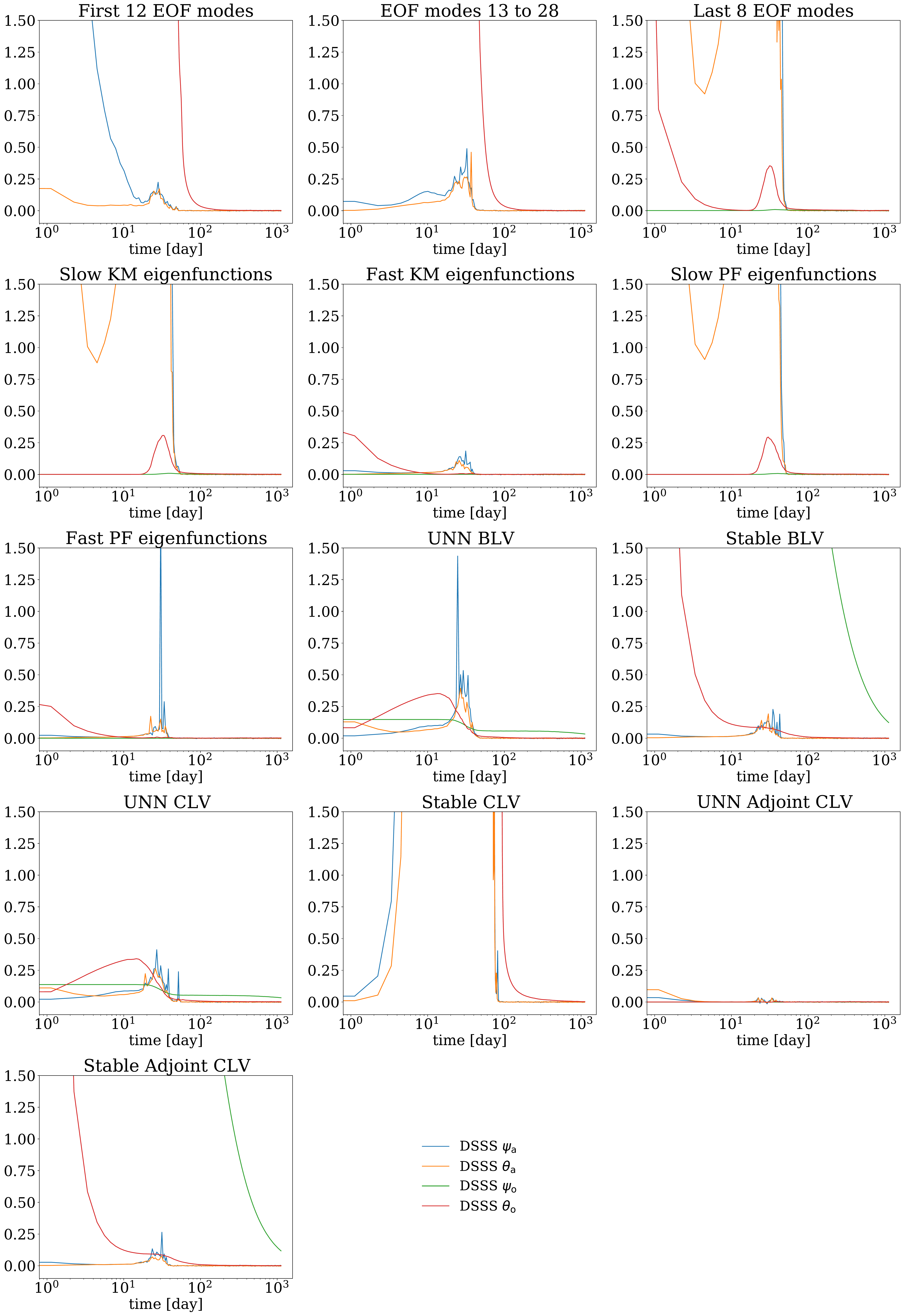}
 \end{center}

 \newpage

\subsection*{Experiment the strong LFV}

\subsubsection*{Case where $\theta_{{\rm o}, 2} >0.12$}

\paragraph*{MSE and Spread as a function of the forecast lead time}

\begin{center}
   \includegraphics[width=\linewidth]{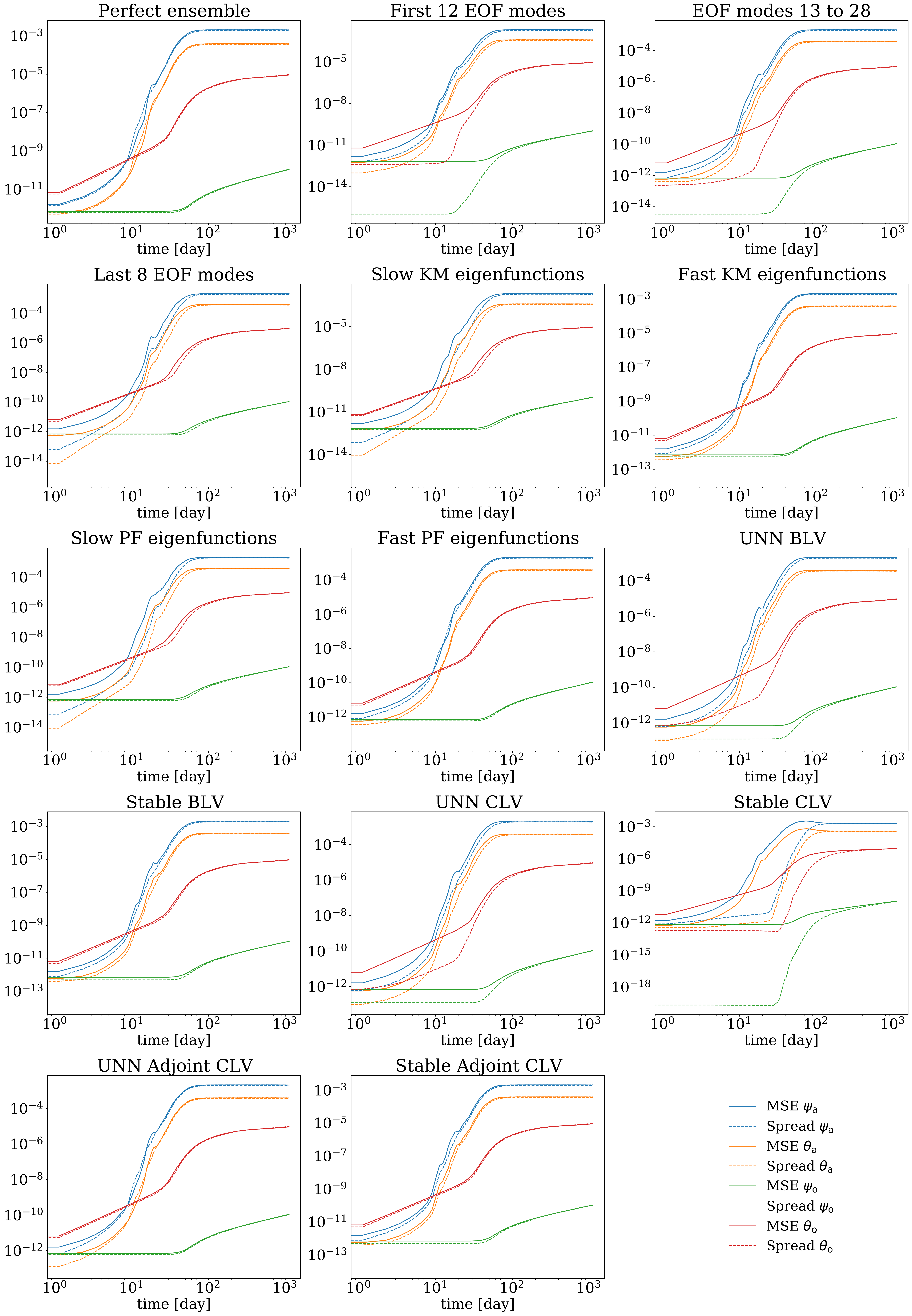}
\end{center}

\newpage

\paragraph*{DSSS as a function of the forecast lead time}

\begin{center}
   \includegraphics[width=\linewidth]{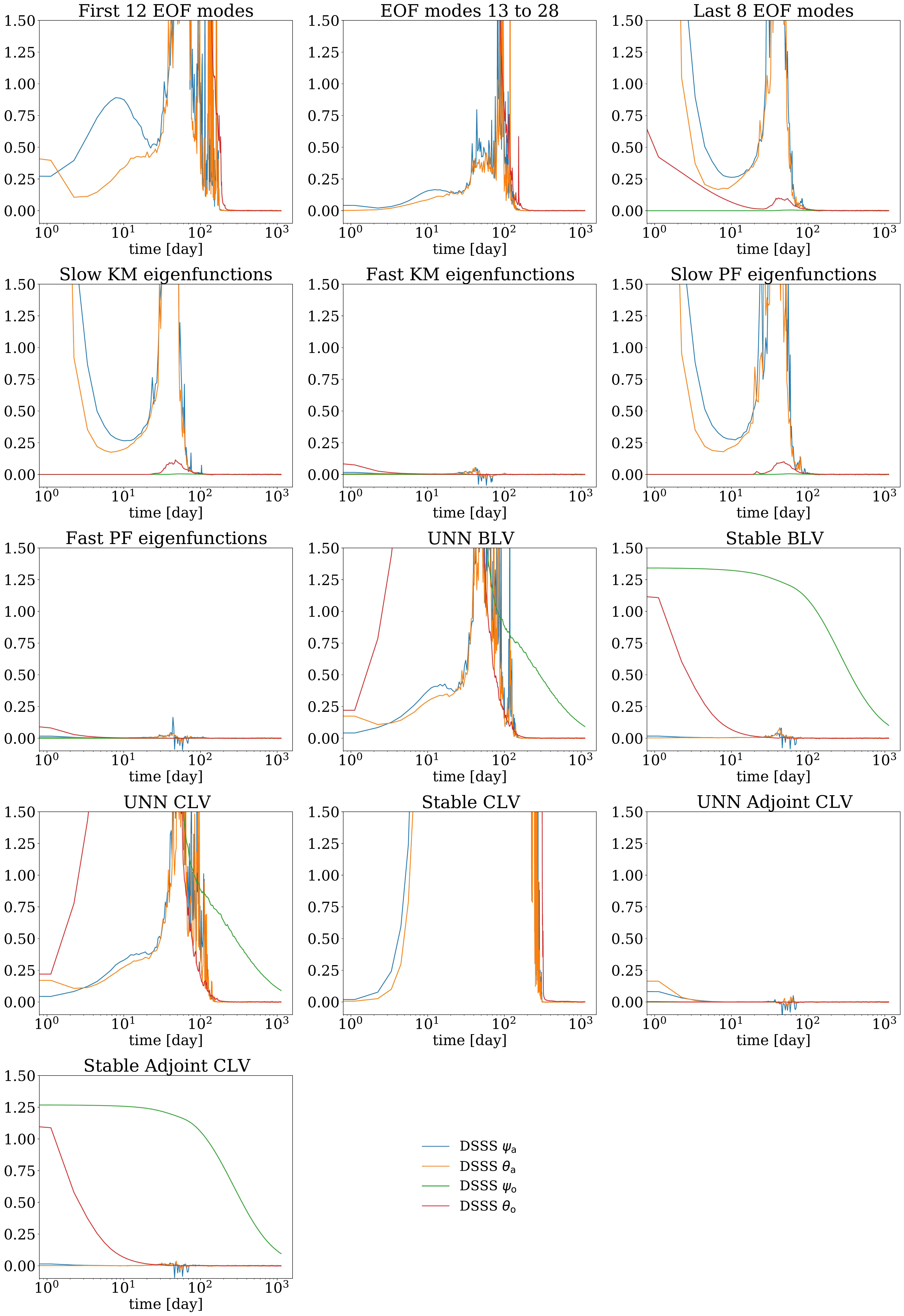}
\end{center}

\newpage

\subsubsection*{Case where $\theta_{{\rm o}, 2} <0.08$}

\paragraph*{MSE and Spread as a function of the forecast lead time}

\begin{center}
   \includegraphics[width=\linewidth]{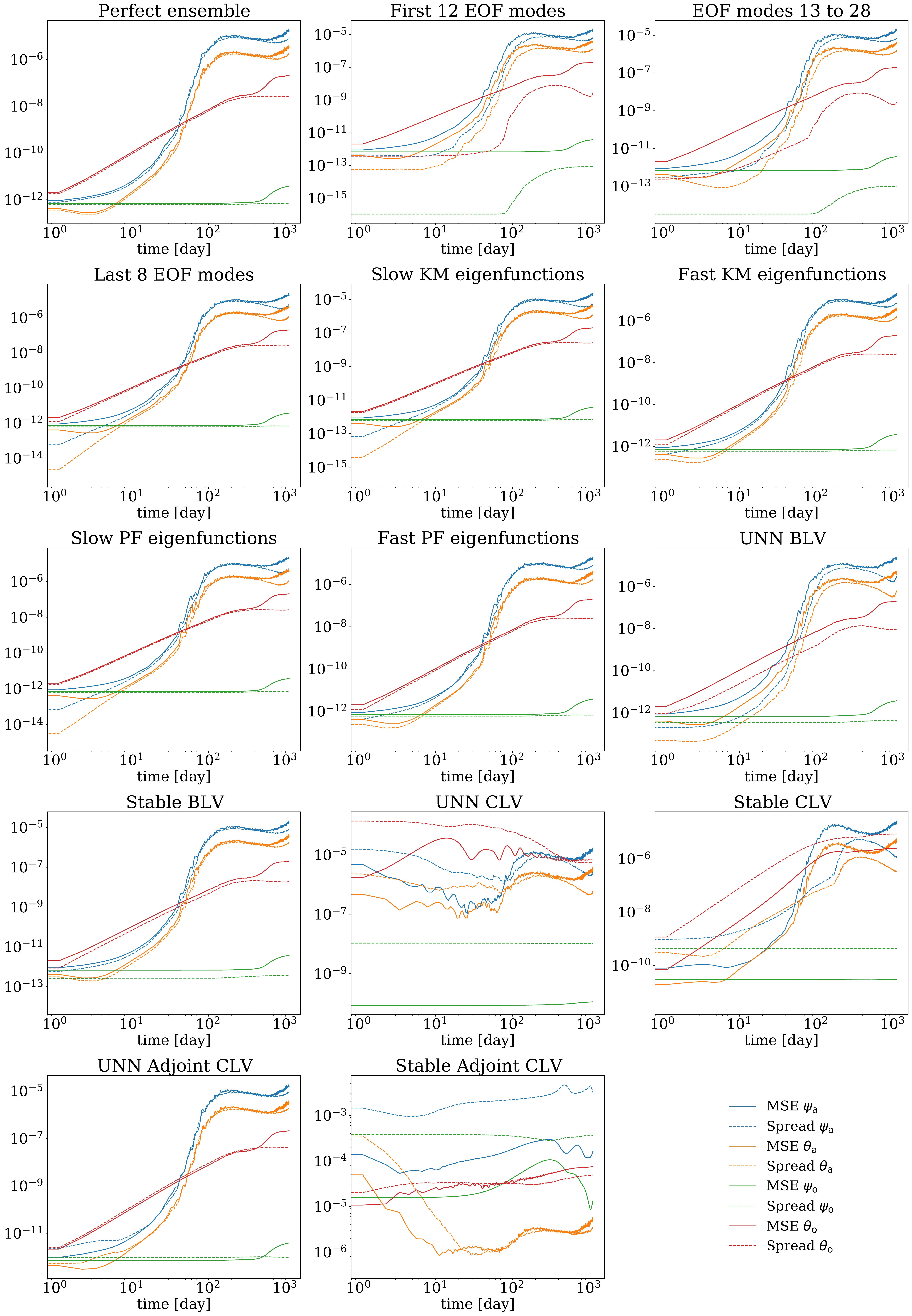}
\end{center}

 \newpage

\paragraph*{DSSS as a function of the forecast lead time}

\begin{center}
   \includegraphics[width=\linewidth]{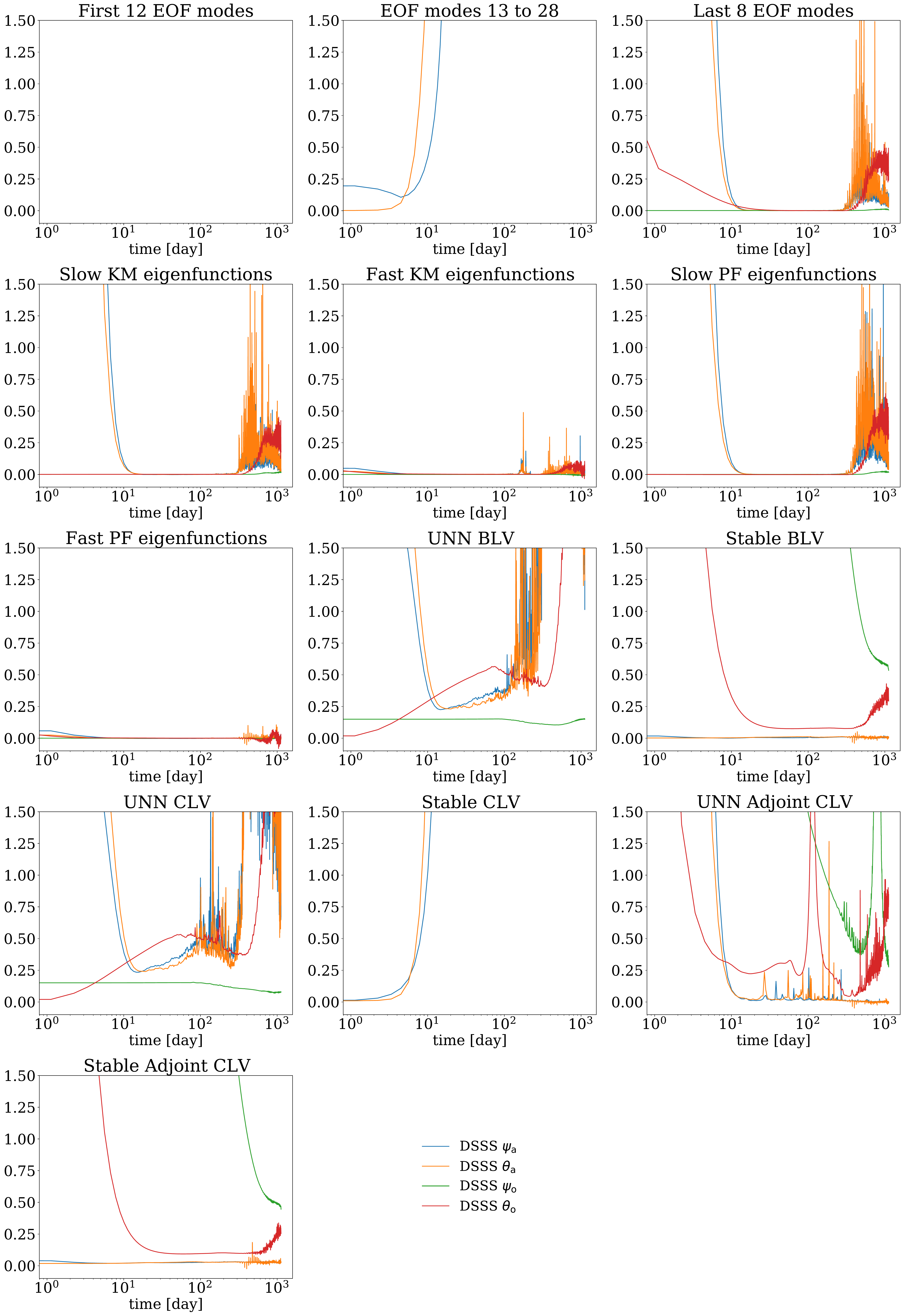}
\end{center}

\newpage

\subsection*{PFMD spectra}

\paragraph*{Experiment without LFV}

\begin{center}
   \includegraphics[width=\linewidth]{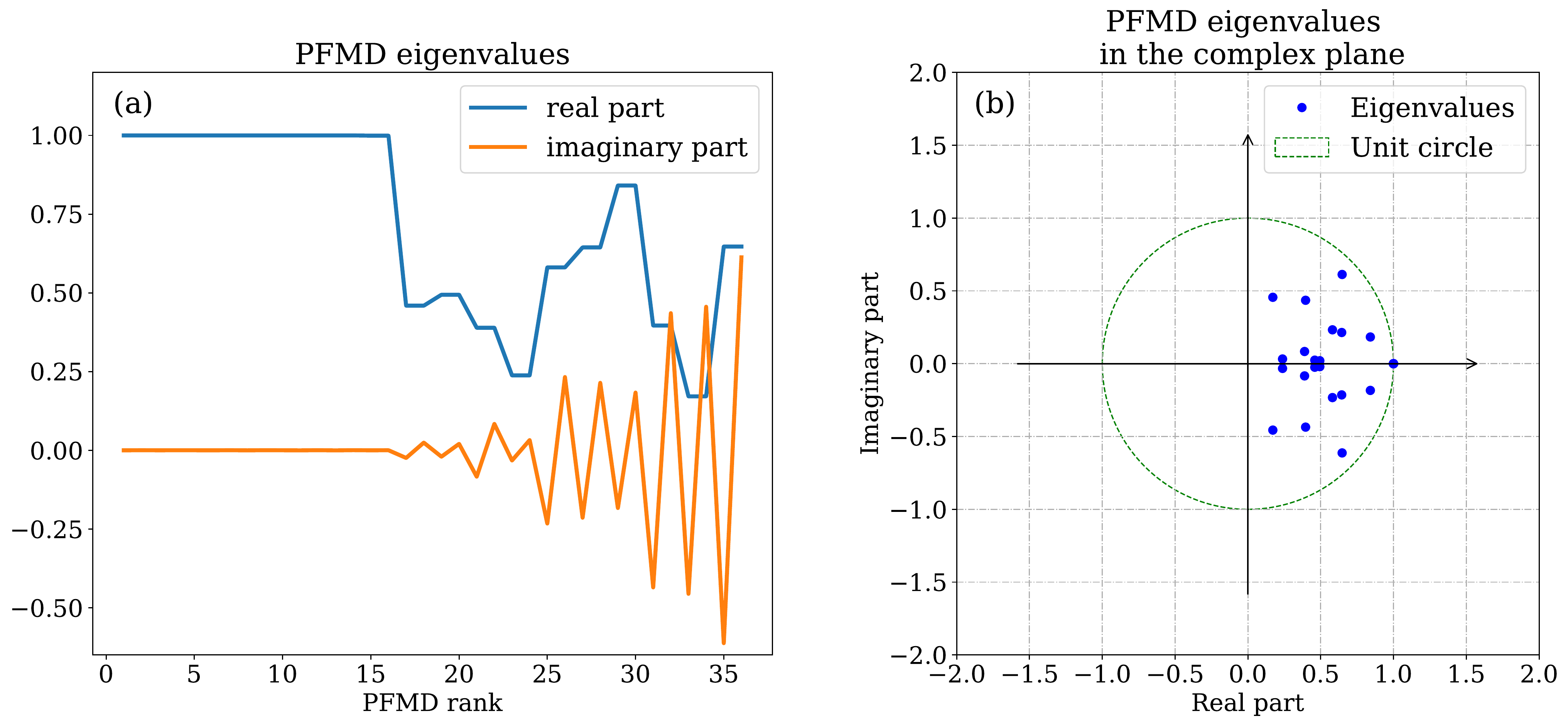}
\end{center}

\paragraph*{Experiment with LFV}

\begin{center}
   \includegraphics[width=\linewidth]{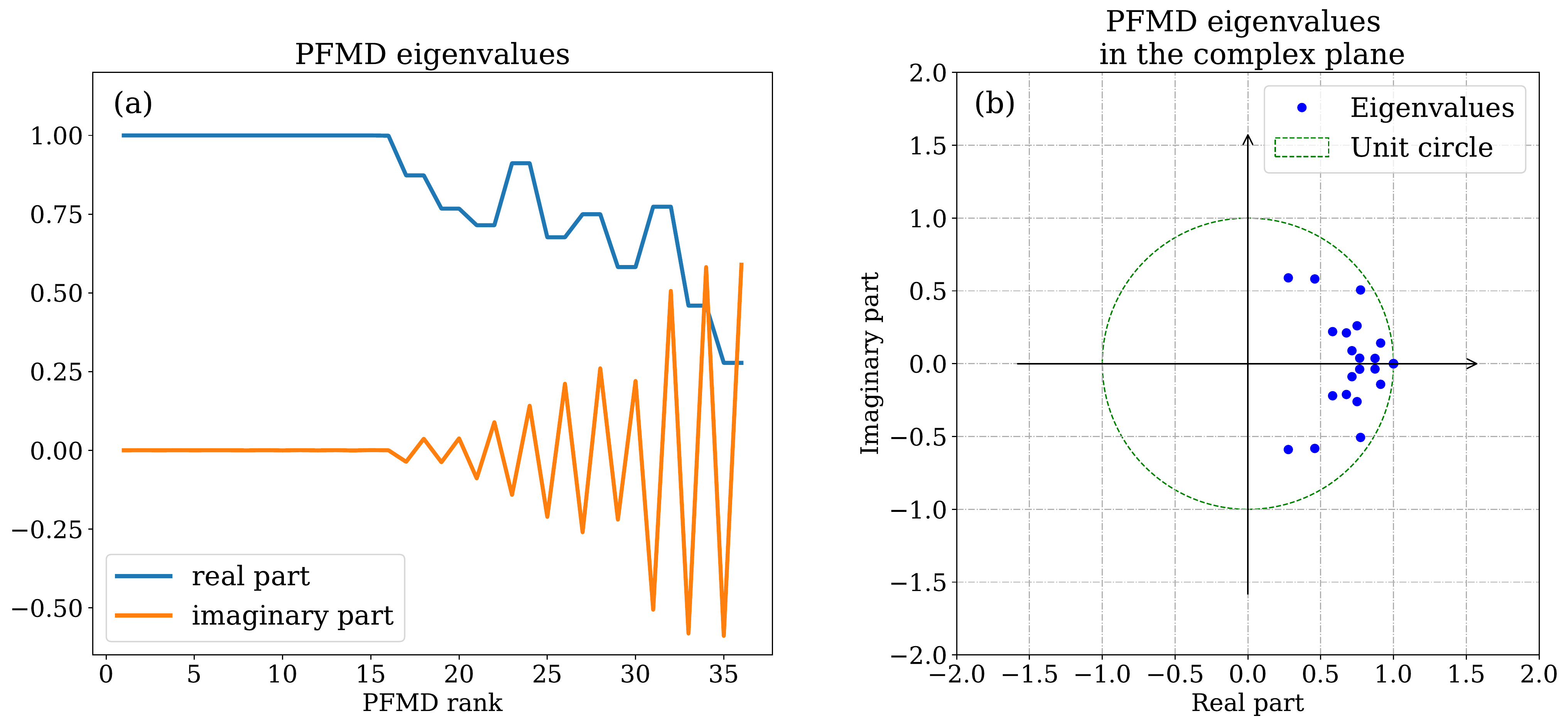}
\end{center}

\end{document}